\documentclass[12pt,showpacs,amsmath,amssymb]{revtex4}
\usepackage{amsmath}

\usepackage{epsfig}
\usepackage{bm}

\newcommand{\Rom}[1]
{\MakeUppercase{\romannumeral #1}}

\begin{document}

\title{Quantum Vacuum: The Structure of Empty Space-Time and Quintessence
with Gauge Symmetry Group $SU(2)\otimes U(1)$ }

\author{Ashot Gevorkyan}
\altaffiliation{$^1$Institute for Informatics and Automation Problems NAS of RA,\\
$^2$Institute of Chemical Physics  NAS of RA }\quad
   \email{g_ashot@sci.am}

\date{\today}
\begin{abstract}
We consider  the formation of structured and massless particles with spin 1,
by using the Yang-Mills like stochastic equations system
for the group symmetry $SU(2)\otimes U(1)$ without taking into
account the nonlinear term characterizing self-action. We prove that,
in the first phase of relaxation, as a result of multi-scale random fluctuations
of quantum fields, massless particles with spin 1, further referred as \emph{hions}, are
generated in the form of statistically stable quantized structures, which are localized
on 2$D$ topological manifolds. We also study the wave state and the geometrical
structure of the \emph{hion} when as a free particle and, accordingly, while
it interacts with a random environment becoming a quasi-particle with a finite
lifetime. In the second phase of relaxation, the vector boson makes
spontaneous transitions to other massless and mass states. The problem of
entanglement of two \emph{hions} with opposite projections of the spins $+1$
and $-1$ and the formation of a scalar zero-spin boson are also thoroughly studied.
We analyze the properties of the scalar field
and show that it corresponds to the Bose-Einstein (BE) condensate.
The scalar boson decay problems, as well as a number of features
characterizing the stability of BE condensate, are also discussed. Then, we report
on the structure of empty space-time in the context of new properties
of the quantum vacuum, implying on the existence of a natural quantum computer
with complicated logic, which manifests in the form of dark energy. The
possibilities of space-time engineering are also discussed.
\end{abstract}

\maketitle

\section{I\lowercase{ntroduction}}
\label{01}
From a mathematical and philosophical point of view, the vacuum can be
comparable to the region of absolutely empty space or, equivalently, with the
region of the space where there are no fields and massive particles (see for
example \cite{Saund}).
One can mention the Lamb Shift \cite{Lamb}, Casimir effect \cite{Casimir},
Unruh effect \cite{Unru}, anomalous magnetic
moment of electron \cite{Proh}, Van der Waals forces \cite{Van}, Delbr\"{u}ck
scattering \cite{Del}, Hawking radiation \cite{Haw}, the cosmological constant
problem  \cite{Weinb,Wein}, vacuum polarization at weak electromagnetic fields
\cite{Ash1,Ash2,Ash3} - here is an incomplete list of phenomena, part of which
has been experimentally discovered and that are  conditioned by the
physical vacuum or, more accurately, by a quantum vacuum (QV).
In reference \cite{Wein} author discuss the issues of cosmic acceleration,
while in \cite{Steinh} dark energy-quintessence are studied  in the framework
of QV  theories, which necessarily include scalar fields. In reference \cite{ENov}
authors evaluate mass and electric dipole moment of the graviton, identifying
as a particle of dark matter, which radically changes our understanding of
dark matter and possibly of dark energy. The properties of QV can be studied in
the scope of quantum field theory (QFT), i.e. quantum electrodynamics (QED) and
quantum chromodynamics (QCD). Note that QFT may accurately describe QV, considering
that one can sum up an infinite series of perturbation theory terms, which is
typical of field theories. However, it is well known that perturbation theory for
QFT is does not converges at low energies, failing to describe, for example, nonzero
values of the vacuum expectation, called condensates in QCD or the BCS superconductivity
theory. In particular, as shown in \cite{Savi},  as well as in \cite{SCHARF},  the radiative
corrections of the massless Yang-Mills theory leads to instability of the vacuum state, which
corresponds to the asymptotic freedom of gauge theories and is due to infrared features.

According to the Standard Model (SM) precisely the non-zero vacuum expectation value
of the Higgs field \cite{Higgs1,Higgs2}, arising from spontaneous symmetry breaking,
is the principle mechanism allowing to generate masses.

 To overcome these difficulties and conduct a consistent and comprehensive QV
study, we propose using complex Langevin-type stochastic differential equations
(SDE) as the basic equation of motion for which Yang-Mills equations serve as the
principle of local correspondence. Note that such mathematical representation
allows to describe the massless quantum fields with multi-scale fluctuations in
Hilbert space, where subspaces with single-particle states of zero mass and
spin 1 exist.

In this study, our goal is to give an unambiguous answer to a number of
important questions of the  QV theory, many of which are still not well understood
or remain open problems. In particular, we believe that it is crucial to answer
to the following questions:
\begin{itemize}
\item Could random fluctuations of massless vector quantum fields lead to the
formation of stable massless  Bose particles with spin 1 -\emph{hions}?
\item What are the space-time features of \emph{hion} and how does its quantum
state change when the multi-scale nature of random fluctuations of massless
fields are taken into account?
\item  What entangled states are created by the two \emph{hions}? How do
 these quantum states evolve on the second phase (scale) of relaxation?
\item What are the properties of the \emph{scalar quantum field-quintessence-dark energy}
consisting of  massless spin-0 bosons?
\item  What is the structure empty space-time taking into account a quantum vacuum? 
\item  Is it possible to implement space-time engineering and, accordingly, change
the fundamental properties of subatomic particles? 
\end{itemize}
 The manuscript is organized as follows. 

In the section \Rom 2, we briefly present some well-known facts about the quantum
motion of a photon, described by the wave function  in the coordinate representation. 

In the section \Rom 3, we justify the multi-scale random nature of the free
Yang-Mills fields. First, we obtain the explicit form of SDE of QVF with the
gauge group symmetry $SU(2)\otimes U(1)$, assuming that all the
self-action terms are identically zero.  Secondly, in the limit of statistical
equilibrium of complex probabilistic processes, the necessary  conditions for
quantization are formulated. Further, the equations of motion for the QVF are derived.
By solving these equations, we obtain a discrete set of orthonormal wave functions
describing the stationary states of a massless Bose particle with spin-1
(\emph{hion}), which are localized on a two-dimensional topological manifold. 

In section \Rom 4 we study in detail the probability distribution in different
quantum states of a \emph{hion}.

In section \Rom 5, we investigate the evolution of the \emph{hion} wave state
on the second relaxation scale.  We show that as a result of a new relaxation,
\emph{hion}  becomes into a quasi-particle, which leads to its spontaneous
transitions to various mass and massless states.

In section \Rom 6, we construct the singlet and triplet states of two entangled
 vector bosons (\emph{hions}). The properties of the scalar bosons (singlet states) and
the possibility of the formation of the Bose-Einstein condensate of these particles
are studied in detail.

In section \Rom 7, we thoroughly discuss the obtained results.

In the section \Rom 8, we  present an important proof confirming the
convergence of the developed theory.

\section{Q\lowercase{uantum motion of a photon in empty space}}
The question of the correspondence between the Maxwell equation and
the equation of quantum mechanics was the focus of attention of many
researchers at the dawn of the development of quantum theory \cite{Opp,Mol,Wein1}.

As shown (see, for example in \cite{BIALYNICKI}), the quantum motion of a
photon in a vacuum can be considered within the framework of a wave function
representation, writing it in vector form similarly to the Weyl equation for
a neutrino:
\begin{eqnarray}
\partial_{t}\bm{\Psi^{\pm}}(\textbf{r},t)\mp c_0\bigl(\textbf{S}\cdot
 \bm\nabla\bigr) \bm{\Psi^{\pm}}(\textbf{r},t)=0,\qquad \partial_{t}\equiv\partial/\partial t,
\label{1.01a}
\end{eqnarray}
where $c_0$ denotes the speed of light in the empty  Minkowski space-time
$(\textbf{r},t)\in \mathbb{R}^4$, with regard to
$\bm{\Psi^{+}}(\textbf{r},t)$ and $\bm{\Psi^{-}}(\textbf{r},t)$, they are the
photon wave functions of both helicities, corresponding to left-handed and
right-handed circular polarizations. In addition, the set of matrices $\textbf{S}
=(S_x,S_y,S_z) $ in a  first-order regular vector equation (\ref{1.01a})
describes an infinitesimal rotations of particles with spin projections $+1$
and $-1$, respectively. These three matrices  can be represented by the form:
\begin{widetext}
\begin{eqnarray}
S_x=\left(
  \begin{array}{ccc}
    0 & 0 & 0\\
    0 & 0 & -i \\
    0 & i & 0 \\
  \end{array}
\right),
\qquad
S_y=\left(
  \begin{array}{ccc}
    0 & 0 & i\\
    0 & 0 & 0 \\
    -i & 0 & 0 \\
  \end{array}
\right),\qquad
S_z=\left(
  \begin{array}{ccc}
    0 & -i & 0\\
    i & 0 & 0 \\
    0 & 0 & 0 \\
  \end{array}
\right).
\label{1.01cy}
\end{eqnarray}
\end{widetext}
Recall that these three matrices are a natural generalization of the Pauli matrices
for $SU(2)$ to $SU(3)$, which formed the basis of the Gell-Mann's quark model \cite{Gell-Mann}.

The absence of electrical and magnetic charges in the equation
(\ref{1.01a}) ensures the following conditions:
\begin{equation}
\nabla \cdot \bm{\Psi^{\pm}}(\textbf{r},t)=0.
\label{1.01}
\end{equation}
If we present the wave function in the form of a Riemann-Zilberstein vector \cite{RieSil}:
\begin{eqnarray}
 \bm{\Psi^{\pm}}(\textbf{r},t)=\frac{1}{\sqrt{2}}\biggl\{\frac{\textbf{D}(\textbf{r},t)}{\sqrt{\epsilon_0}}
 \pm i\frac{\textbf{B}(\textbf{r},t)}{\sqrt{\mu_0}}\biggr\},\qquad c_0=\frac{1}{\sqrt{\mu_0\epsilon_0}},
\label{1.01b}
\end{eqnarray}
then from the Eqs. (\ref{1.01a}) and (\ref{1.01}) it is easy to find Maxwell's equations
in an ordinary vacuum or in empty space:
\begin{eqnarray}
\partial_t \textbf{D} -{\bm\nabla}\times\textbf{H} =0,\qquad
 \nabla\cdot\textbf{E}=0,\nonumber\\
\partial_t \textbf{B} +\,{\bm\nabla}\times\textbf{E}=0,\qquad
 \nabla\cdot\textbf{H}=0,
 \label{1.01c}
\end{eqnarray}
where $\epsilon_0$ and $\mu_0$ describe the electric and magnetic constants of
the vacuum, respectively. It is important to note that the dielectric and magnetic
constants provide the following equations:
$$\textbf{D}=\epsilon_0\textbf{E}, \qquad \textbf{B}=\mu_0\textbf{H}.$$
Recall that the only difference between the equations (\ref{1.01a}) and (\ref{1.01c})
is that the Maxwell equations system does not explicitly take into account the spin
of the photon, which is important for further theoretical constructions.

Since  $\epsilon_0$ and $\mu_0$ are constants independent of external
fields and characterizing the state of a free or \emph{unperturbed vacuum},
an idea arises: to consider a vacuum or, more precisely, QV, as some
energy environment with \emph{unusual properties and structures}.

\section{V\lowercase{ector field and its fundamental particle-\underline{\emph{hion}}}}

\subsection{Yang-Mills theory for free fields}
The Yang-Mills theory is a special example of gauge field theory
with a non-Abelian gauge symmetry group, whose Lagrangian for the free-fields
case has the following form (see \cite{Yang}, as well as \cite{Caprini}):
\begin{eqnarray}
\mathcal{L}_{gf}=-\frac{1}{2}\mathrm{Tr}(\mathcal{F}^2)=-
\frac{1}{4}\mathcal{F}^{\mu\nu}_{a}\mathcal{F}^a_{\mu\nu},
\label{2.l1}
\end{eqnarray}
where $\mathcal{F}$ is the 2-form of the Yang-Mills field strength.\\
Note that it remains invariant under impact of the tensor potential
$A_{\mu}^a$ of the gauge group is affected:
\begin{eqnarray}
\mathcal{F}^a_{\mu\nu}=\partial_\mu \mathcal{A}^a_\nu-\partial_\nu\mathcal{A}^a_\mu
+\mathfrak{g}\mathfrak{f}^{abc}\mathcal{A}^b_\mu \mathcal{A}^c_\nu,\qquad
\mu,\nu=0,1,2,3,
\label{2.l2}
\end{eqnarray}
where  $\partial_\mu=(ic_0^{-1}\partial_t,\partial_x,\partial_y,\partial_z)=
(ic_0^{-1}\partial_t,{\bm \nabla})$ denotes the covariant derivative in the
four-dimensional Minkowski space-time, which
in Galilean coordinates is reduced to the usual partial derivative. In addition,
$\mathfrak{f}^{abc}=\mathfrak{f}_{abc}$ are called structural constants of the
group (Lie algebra), $\mathfrak{g}$ is the interaction constant and, finally,
for the group $SU(N)$, the number of isospins  generators varies
within $a,b,c =[1, N^2-1].$

From the  Lagrangian (\ref{2.l1}) one can derive the equations of motion
for the classical free Yang-Mills fields:
\begin{eqnarray}
\partial^\mu \mathcal{F}^a_{\mu\nu}+\mathfrak{g}\mathfrak{f}^{abc}
\mathcal{A}^{\mu b}\mathcal{F}^c_{\mu\nu}=0,
\label{2.l3}
\end{eqnarray}
which are obviously characterized by self-action. Note that in the case of a small
coupling constant $\mathfrak{g} <1$, the perturbation theory is applicable for
solving these equations. However, as shown by numerous studies, in the case
$\mathfrak{g}<1$, massless vector bosons with spin 1 are not formed in the
free Yang-Mills fields. In other words, it remains to assume that the coupling
constant should be greater than $\mathfrak{g}>1$, but then it is not clear how
to solve the equations (\ref{2.l3}) and, accordingly, the problem remains open \cite{Cly}.

Note that, as shown by numerous studies, even with relatively weak nonlinearity,
the behavior of Yang-Mills fields is chaotic in large regions of the phase space
(see, for example, \cite{Matinyan}).  However, the chaotic behavior of the Yang-Mills
fields, in our opinion, may have a different, no less fundamental nature associated
with the global behavior of quintessence. In particular, as follows from various
experiments, on very small space-time scales, continuous fluctuations of vacuum
fields become so significant that the inclusion of these contributions to the basic
equations of motion becomes a fundamentally necessary task. In other words, it
would be quite natural if we assumed that the Lagrangian (\ref{2.l1}) or more precisely,
the vacuum fields $\mathcal{F}^a_{\mu\nu} $ are random. The latter is obviously
equivalent to the requirement that the basic equations of motion (\ref{2.l3}) be
\emph{stochastic differential equations} (SDE). In this regard, the natural
question arises: how to quantize these fields?  It is obvious that canonical
quantization, i.e.  the functional integral methods for a generating functional
the $n$-point functions, cannot be used in this case.

It would seem that in order to solve the problem of field quantization, the method
of stochastic quantization Parisi-Wu \cite{Dam} is more adapted, which is based on
an important analogy between Euclidean quantum field theory and classical statistical
mechanics. Without going into details, we note that the Paris-Wu approach, despite
 a number of serious advantages over other approaches, in particular, the absence
 of problems with the Faddeev-Popov ghost fields \cite{Fad-Pop}, is problematic to
 apply to such a critical substance as quantum vacuum.

\subsection{Q\lowercase{uantization of stochastic vacuum fields}}
Let us consider the simplest case, where space-time is described by
the Lorentz metric $\chi_{\mu\nu}=\mathrm{diag}(+\,-\,-\,-)$, the coupling
constant $\mathfrak{g}=0$, and when the fields satisfy  the symmetry group
$SU(2)\otimes U(1)$. The latter means that we consider the unified electroweak
interaction within the framework of the Abelian gauge group, but using stochastic
field equations.  It should, however, be noted that in the case when
$\mathfrak{g}=0$, the isospins do not interact with each other. 

We determine the covariant antisymmetric tensor of the  quantum  vacuum fields 
(QVF) in the form:
\begin{equation}
\mathcal{F}^a_{\mu\nu}=
\left(
  \begin{array}{cccc}
0 & \psi_x^\pm & \psi_y^\pm & \psi_z^\pm\\
\psi_x^\pm & 0 &\pm\psi_z^\pm &\mp \psi_y^\pm\\
\psi_y^\pm & \mp \psi_z^\pm & 0&\mp \psi_x^\pm\\
\psi_z^\pm &\pm \psi_y^\pm &\mp \psi_x^\pm & 0\\
  \end{array}
\right).
\label{2.l0}
\end{equation}

 In the case when $\mathfrak{g} = 0 $, it seems logical to refine the equation
(\ref{2.l3}) for short distances  (see for example \cite{Faizal}), which leads to the equation:
$$\partial^\mu \mathcal{F}^a_{\mu\nu}-\frac{\alpha}{15\pi m^2}\square\partial^\mu
\mathcal{F}^a_{\mu\nu}=0,$$ where $\square=\triangle-c^{-2}\partial_t^2$ denotes the D'Alembert operator,
$\bigtriangleup$ is the Laplace operator, $\alpha$ is the fine structure constant
and $m$ is the mass of an electron. However, if to make the substitution
$\tilde{\mathcal{F}}^a_{\mu\nu}=\mathcal{F}^a_{\mu\nu}-\frac{\alpha}{15\pi m^2}\square
\mathcal{F}^a_{\mu\nu}$, then we get the equation $\partial^\mu\tilde{\mathcal{F}}^a_{\mu\nu}=0.$
Below we will study the properties of generalized QVF $\tilde{\mathcal{F}}^a_{\mu\nu}$. 

 Since for $\mathfrak{g}=0$  the equations (\ref{2.l3})
of all isospins are symmetric, below, where it does not cause confusion, the index
$a=(1,2,3)$ will be omitted. 

 Now we will consider the equation (\ref{2.l3}), taking into account the fact that
$\mathfrak{g}=0$.\\ Substituting (\ref{2.l0}) into the equation (\ref{2.l3}), one can
obtain the following three independent SDEs: 
\begin{eqnarray}
ic^{-1}\dot{\psi}^{\pm}_x=\pm\partial_y \psi^{\pm}_z\mp\partial_z \psi^{\pm}_y,
\nonumber\\
ic^{-1}\dot{\psi}^{\pm}_y=\pm\partial_z\psi^{\pm}_x\mp\partial_x \psi^{\pm}_z,
\nonumber\\
ic^{-1}\dot{\psi}^{\pm}_z=\pm\partial_x\psi^{\pm}_y\mp\partial_y \psi^{\pm}_x,
\label{1.02tab}
\end{eqnarray}
and the following equation between of derivatives of the field components:
\begin{eqnarray}
 \partial_x{\psi}^{\pm}_x+ \partial_y{\psi}^{\pm}_y+\partial_z{\psi}^{\pm}_z=0,
 \label{1.02ta}
\end{eqnarray}
where $c$ is the velocity of the field propagation, which may differ
from the speed of light in \emph{ordinary vacuum}, in addition,
$\dot{{\psi}}^{\pm}_\sigma= \partial{\psi}^{\pm}_\sigma/\partial t$
and $\sigma=x,y,z$.

Combining SDE (\ref{1.02tab}), we can write the following  stochastic  vector equation:
\begin{eqnarray}
 \dot{\bm{\psi}}^{\pm}\bigl(\textbf{r},t;\textbf{f}(t)\bigr) \mp
 c\bigl(\textbf{S}\cdot \bm\nabla\bigr)
\bm{\psi^{\pm}}\bigl(\textbf{r},t;\textbf{f}(t)\bigr) =0,\qquad t\in(-\infty,+\infty),
\label{2.0a1}
\end{eqnarray}
where $\textbf{f}(t)$ is a random function, which is clearly defined below (see Eqs.
(\ref{1.0k2a}) and (\ref{1.0t2a})),  and, accordingly, the function
$\bm{\psi^{\pm}}\bigl(\textbf{r},t;\textbf{f}(t)\bigr)$, in this case will
mathematically have the sense of a complex probabilistic process (see for example
\cite{AshG,Ashg1}), which can be represented as a three-component vector in
a Hilbert space: 
\begin{eqnarray}
\label{1.01cz}
\bm\psi^{\pm}\bigl(\textbf{r},t;\textbf{f}(t)\bigr)=\left[
\begin{array}{ccc}
\psi^{\pm}_x\bigl(\textbf{r},t;\textbf{f}(t)\bigr)\\
\psi^{\pm}_y\bigl(\textbf{r},t;\textbf{f}(t)\bigr)\\
\psi^{\pm}_z\bigl(\textbf{r},t;\textbf{f}(t)\bigr)\\
\end{array}
\right].
\end{eqnarray}
Note that $\bm{\psi^{\pm}}\bigl(\textbf{r},t;\textbf{f}(t)\bigr)\in
L^2(\mathbb{R}^4\otimes \textbf{R}_{\{\textbf{f}\}}),$ where
$\textbf{R}_{\{\textbf{f}\}}$ denotes the functional space. Recall that
in the case when $\textbf{f}(t)\equiv0$, the equation (\ref{2.0a1}) goes into the
usual Weyl equation of type (\ref{1.01a})-(\ref{1.01cy}).

As for the symmetry of the electroweak fields, they satisfy the
following commutation relations:
\begin{equation}
 [T_a,T_b]=i\sum_{c=1}^3\mathfrak{f}_{abc}T_c,\qquad T_a=\frac{1}{2}S_a,
 \label{1.l2k}
\end{equation}
where $S_1=S_x,\,S_2=S_y$ and $S_3=S_z$ (see  expressions  (\ref{1.01cy})).\\
Note that the structural constants are subject to the following relations;
$\mathfrak{f}_{123}=-\mathfrak{f}_{132}=\mathfrak{f}_{312}
=-\mathfrak{f}_{321}=\mathfrak{f}_{231}=1/2$ and ${\emph{Tr}}\,(T_a)=0$.

It is well known that quantum vacuum fields are characterized by random multi-scale
fluctuations in four-dimensional Minkowski space-time. 
The latter, as a result, leads to a multi-scale evolution of these fields.
Note that, in simplest case the multi-scale evolution of the system can be
characterized by two sets of parameters $\{\tau,\bm\varepsilon\}$, by the
relaxation times $\{\tau\}=(\tau_0,\tau_{1}, ...)$ and fluctuation
powers $\{\bm\varepsilon\}=(\bm{\varepsilon}_0,\bm{\varepsilon}_1, ...,) $,
respectively. It is also assumed that at small time intervals $\delta{t}<<\tau_0$,
where $\tau_0=min\{\tau\}$ is the relaxation time of the minimum duration, the
complex SDE (\ref{2.0a1}) passes to the Weyl type equation (\ref{1.01a})-(\ref{1.01cy}).
In other words, the equation (\ref{1.01a})-(\ref{1.01cy}) in this case plays the role of
the principle of local correspondence, and therefore further the stochastic equation
(\ref{2.0a1}) will conventionally be called the \emph{Langevin-Weyl} equation.

Note that our main goal will be to study the equation (\ref{2.0a1}) for the
symmetry group $SU(2)\otimes U(1)$ on the main relaxation scale, characterized by
parameters $(\tau_0,\bm{\varepsilon}_0)$. Recall that for each symmetry group these
parameters are different. In particular, for the symmetry under consideration,
multi-scale fluctuations should obviously be characterized by three constants
$({\bm\varepsilon}_{0}^1,{\bm\varepsilon}_{0}^2,{\bm\varepsilon}_{0}^3)$
(see the definition of the number of isospin generators after the equation
(\ref{2.l2}) and, respectively, below after the equation (\ref{1.0t2a})).

\textbf{Theorem.} \emph{If QVF obeys the Langevin-Weyl SDE (\ref{2.0a1}), then
for the symmetry group $SU(2)\otimes U(1)$ on the main relaxation scale
$(\tau_0,\bm{\varepsilon}_0^a)$, in the limit of statistical equilibrium, a massless
Bose particle with spin 1 is formed as a 2D topological structure in 3D space.} 

Obviously, in the case of a localized quantum state, the four-dimensional
interval of the propagated signal will be equal to zero, and, respectively,
the points of the Minkowski space (events) must be connected by a relation
similar to the light cone:
\begin{equation}
s^2 =c^2t^2-r^2=0, \qquad r^2=x^2+y^2+z^2.
 \label{1.02k}
\end{equation}
Bearing in mind that particles with projections of spins $+1$ and $-1$ are
symmetric, below we will investigate only the wave function of a particle
with spin projection +1.

Taking into account (\ref{1.02tab}) and (\ref{1.02ta}), we obtain the following
second-order partial differential equations:
\begin{eqnarray}
\square \psi^{+}_x={c}^{-1}{c_{,y}}\bigl(\partial_x \psi^{+}_y-\partial_y \psi^{+}_x\bigr)-
{c}^{-1} {c_{,z}} \bigl(\partial_z \psi^{+}_x-\partial_x \psi^{+}_z \bigr)- c_{,t}c^{-3}\dot{\psi}^{+}_x,
 \nonumber\\
\square\psi^{+}_y={c}^{-1}{c_{,z}}\bigl(\partial_y \psi^{+}_z-\partial_z \psi^{+}_y\bigr)-
{c}^{-1}{c_{,x}}\bigl(\partial_x \psi^{+}_y-\partial_y \psi^{+}_x\bigr)-c_{,t}c^{-3}\dot{\psi}_y,
\nonumber\\
\square\psi^{+}_z= {c}^{-1}{c_{,x}}\bigl(\partial_z \psi^{+}_x-\partial_x \psi^{+}_z\bigr)-
{c}^{-1}{c_{,y}}\bigl(\partial_y \psi^{+}_z-\partial_z \psi^{+}_y\bigr)-c_{,t}c^{-3}\dot{\psi}_z,
\label{1.02t}
\end{eqnarray}
where  $c_{,\breve{\sigma}}=\partial c/\partial\breve{\sigma}$ and $\breve{\sigma}=(x,y,z,t)$.

Now we need to determine the explicit form of the equations for this it is
necessary to calculate the derivatives $c_{,\breve {\sigma}}$.
Using the equations (\ref{1.02k}), it is easy to calculate:
\begin{equation}
c_{,t}=- c^2/r,\qquad c_{,x}= cx/r^{2}, \qquad c_{,y}= cy/r^{2}, \qquad
 c_{,z}= cz/r^{2}.
\label{1.02zt}
\end{equation}
For further analytic study of the problem, it is useful to bring the system of equations
(\ref{1.02t}) into  canonical form, when all components of the field are separated
and each of them is described by a separate equation.

In particular, by courting the fact that in the problem under consideration all fields
are symmetric, the following additional conditions can be imposed on the field components:
\begin{eqnarray}
 \bigl(c_{,z}- c_{,y}\bigr)\dot{\psi}_x=c_{,z}\dot{\psi}_y^+-c_{,y}\dot{\psi}_z^+,
\nonumber\\
\bigl(c_{,x}- c_{,z}\bigr)\dot{\psi}_y=c_{,x}\dot{\psi}_z^+-c_{,z}\dot{\psi}_x^+,
\nonumber\\
\bigl(c_{,y}- c_{,x}\bigr)\dot{\psi}_z=c_{,y}\dot{\psi}_x^+-c_{,x}\dot{\psi}_y^+.
 \label{1.02ak}
\end{eqnarray}
It is easy to verify that these conditions are symmetric with respect to
the components of the field and are given on the hyper-surface of four-dimensional
events. Using the conditions (\ref{1.02ak}), the system of equations (\ref{1.02t})
can be easily reduced to the canonical form:
\begin{eqnarray}
\Bigl\{\square+\bigl[i(c_{,z}-c_{,y})+c_{,t}c^{-1}\bigr]c^{-2}\partial_t\Bigr\}\psi^{+}_x=0,
\nonumber\\
\Bigl\{\square+\bigl[i(c_{,x}-c_{,z})+c_{,t}c^{-1}\bigr]c^{-2}\partial_t\Bigr\}\psi^{+}_y=0,
\nonumber\\
\Bigl\{\square+\bigl[i(c_{,y}-c_{,x})+c_{,t}c^{-1}\bigr]c^{-2}\partial_t\Bigr\}\psi^{+}_z=0.
 \label{1.02zkl}
\end{eqnarray}

For further investigations, it is convenient to represent the wave function
component in the form:
 \begin{equation}
 \psi_\sigma^+\bigl(\textbf{r},t;f_\sigma(t)\bigr)=\exp\biggl
 \{\int^t_{-\infty} \zeta_\sigma(t')dt'\biggr\}\phi_\sigma^+(\textbf{r}),
 \label{1.02a}
\end{equation}
where $\zeta_\sigma(t)$ denotes the random function, and $f_\sigma(t)$ is
the corresponding projection of the random vector $\textbf{f}(t)$.

Substituting (\ref{1.02a}) into (\ref{1.02zkl}) and taking into account (\ref{1.02zt}),
we get the following system of differential equations:
\begin{eqnarray}
\biggl\{\triangle-\Bigl[\Bigl(\frac{\xi_x(t)}{c}\Bigr)^2+\frac{r-i(z-y)}{cr^2}
 \zeta_x(t)\Bigr]\biggr\}\phi_x^+(\textbf{r})=0,\,\,
\nonumber\\
\biggl\{\triangle-\Bigl[\Bigl(\frac{\xi_y(t)}{c}\Bigr)^2+\frac{r-i(x-z)}{c r^2}
\zeta_y(t)\Bigr]\biggr\}\phi_y^+(\textbf{r})
=0,\,\,
\nonumber\\
\biggl\{\triangle-\Bigl[\Bigl(\frac{\xi_z(t)}{c}\Bigr)^2+\frac{r-i(y-x)}{cr^2}
\zeta_z(t)\Bigr]\biggr\}\phi_z^+(\textbf{r})=0.\,\,
\label{1.02b}
\end{eqnarray}
In the equations (\ref{1.02b}) the following notation is made:
$$\xi^2_\sigma(t)=\dot{\zeta}_\sigma(t)+\zeta^2_\sigma(t),$$
where $\dot{\zeta}_\sigma(t)=\partial\zeta_\sigma(t)/\partial t$.

It is easy to verify that the coefficients in the equations (\ref{1.02b}),
are random functions of time. It will be natural if we average these
equations on the scale of the relaxation time $\tau_0$.

Averaging the equations (\ref{1.02b}), we get the following system of
second-order stationary differential equations:
\begin{eqnarray}
\biggl\{\triangle-\Bigl[\Bigl(\frac{\omega_x}{c}\Bigr)^2+\frac{r-i(z-y)}{cr^2}
\varrho(\omega_x)\Bigr]\biggr\}\phi_x^+(\textbf{r})=0,
\nonumber\\
\biggl\{\triangle-\Bigl[\Bigl(\frac{\omega_y}{c}\Bigr)^2+\frac{r-i(x-z)}{c r^2}
\varrho(\omega_y)\Bigr]\biggr\}\phi_y^+(\textbf{r})=0,
\nonumber\\
\biggl\{\triangle-\Bigl[\Bigl(\frac{\omega_z}{c}\Bigr)^2+\frac{r-i(y-x)}{cr^2}
\varrho(\omega_z)\Bigr]\biggr\}\phi_z^+(\textbf{r})=0,
\label{2.0k2b}
\end{eqnarray}
where $\omega_\sigma$ and $\varrho(\omega_\sigma)$ are regular parameters of the
problem, which are defined as follows:
\begin{equation}
\omega^2_\sigma=\langle\xi^2_\sigma(t)\rangle=\langle\dot{\zeta}_\sigma(t)+
\zeta^2_\sigma(t)\rangle,\qquad \varrho(\omega_\sigma)=\langle\zeta_\sigma(t)\rangle.
\label{2.0k2a}
\end{equation}
In the (\ref{2.0k2a}) the bracket $\langle...\rangle$ denotes the averaging
operation by the relaxation time $\tau_0$.

Now the main question is that: is it possible the emergence of statistical
equilibrium in the system under consideration, which can lead to the stable
distribution of the parameter $\varrho(\omega_\sigma)$?

Note that the latter circumstance, for obvious reasons, eliminates the nontrivial
question connected with the unitary transformation of the state vector, since
the quantum system in this problem is not isolated. Obviously, in this case
it is necessary to require the conservation of the norm of the average
value of the state vector:
\begin{eqnarray}
\bigl\langle\bm\psi^{\pm}\bigl(\textbf{r},t;\textbf{f}(t)\bigr)
\bigr\rangle_{\textbf{R}_{\{\textbf{f}\}}}= \bm\phi^{\pm}(\textbf{r})=\left[
\begin{array}{ccc}
\phi^{\pm}_x(\textbf{r})\\
\phi^{\pm}_y(\textbf{r})\\
\phi^{\pm}_z(\textbf{r})\\
\end{array}
\right],
\label{2.07c}
\end{eqnarray}
where
$$
\phi^{\pm}_\sigma(\textbf{r})=\bigl\langle\psi^{\pm}_\sigma\bigl(\textbf{r},t;
f_\sigma(t)\bigr)\bigr\rangle_{R^\sigma_{\{f_\sigma\}}},\qquad\sigma=x,y,z,
$$
the bracket $\langle...\rangle_{R^\sigma_{\{f\}}}$ denotes the functional integration:
$$
\textbf{R}_{\{\textbf{f}\}}=R^x_{\{f_\sigma\}}\otimes R^y_{\{f_\sigma\}}\otimes
R^z_{\{f_\sigma\}}.
$$
Now the key question for the representation will be the proof of
existence of an average value of the state vector $\bm\phi^{\pm}(\textbf{r})$
in the limit of statistical equilibrium or formally at $t\to\infty$.

Using the first relation in (\ref{2.0k2a}), we can define the following Langevin
equation:
\begin{eqnarray}
\dot{\zeta}_\sigma=-(\zeta^2_\sigma-\omega^2_\sigma)+U_\sigma(t),
\label{1.0k2a}
\end{eqnarray}
where $U_\sigma(t)=U_{0 \sigma}+f_\sigma(t),$ in addition,
$U_{0 \sigma}=\langle{U_\sigma(t)}\rangle<0$ is an unknown constant. As for the
term $f_\sigma(t)$,  it denotes a random force that satisfies the \emph{white noise}
correlation relations:
\begin{equation}
\langle f_\sigma(t)\rangle=0,\qquad \langle{f_\sigma(t)f_\sigma(t')}\rangle
=\varepsilon_{0\sigma}^a\delta(t-t').
\label{1.0t2a}
\end{equation}
 Recall that in (\ref{1.0t2a}) the set of constants
${\bm\varepsilon}_{0}^a=(\varepsilon_{0x}^a,\varepsilon_{0y}^a,\varepsilon_{0z}^a)$
denote the oscillation powers of isospin $a$ along different axes.
It is natural to assume that for each isospin ${\varepsilon}_{0}^a=\varepsilon_{0x}^a=
\varepsilon_{0y}^a=\varepsilon_{0z}^a$, whereas
  $\varepsilon_{0y}^a\neq\varepsilon_{0z}^b$  when $a\neq b$. 

Recall that in $SU(2)\times U(1)$ gauge group there are three $W$ bosons
of weak isospin from $SU(2)$, namely ($W_1,W_2,$ and $W_3$), and the $B$
boson of weak hypercharge from $U(1)$, respectively, all of which are massless.
These bosons obviously must be characterized by a set of constants
$({\varepsilon}_0^1,{\varepsilon}_0^2,{\varepsilon}_0^3)$.  

Using SDE (\ref{1.0k2a}) and relations (\ref{1.0t2a}), as well as assuming that
$U_{0\sigma}=-2\,\omega^2_\sigma$, one can obtain the equation for the probability distribution
\cite{Klyat,Lif}):
\begin{equation}
\frac{\partial\mathcal{P}^0}{\partial t}=\biggl\{\frac{\partial}{\partial \zeta}
\bigl(\zeta^2+\omega^2\bigl)
+\frac{\varepsilon_{0}}{2}\frac{\partial^2}{\partial\zeta^2}\biggr\}\mathcal{P}^0.
\label{2.0t2a}
\end{equation}
Recall that in (\ref{2.0t2a}) and below, to simplify writing, we will omit both
the isospin index $a$ and the index $\sigma$, which denotes the coordinate. 

Solving the equation (\ref{2.0t2a}), it is easy to find \cite{Frish}:
\begin{equation}
\mathcal{P}^0(\bar{\zeta};\bar{\omega})=2\varepsilon^{-1}_{0}\mathcal{J}
(\bar{\omega})e^{-2\Phi(\bar{\zeta})}
\int^{\bar{\zeta}}_{-\infty}e^{2\Phi(\bar{\zeta}')}d\bar{\zeta}',
\label{2.32a}
\end{equation}
where $\bar{\zeta}=\zeta/\varepsilon^{1/3}_0$ and $\Phi(\bar{\zeta})=
(\bar{\zeta}^3+3\bar{\omega}^2\bar{\zeta})/3$.\\
As for the coefficient $\mathcal{J}(\bar{\omega})$, it is determined from
the normalization condition of the distribution (\ref{2.32a}) to the unity:
$$
\mathcal{J}^{-1}(\bar{\omega})=\sqrt{\pi}\biggl(\frac{2}{\varepsilon_0}\biggr)^{1/3}
\int_{0}^{\infty}
\exp\Bigl[-\frac{x^3}{6}-2\,\bar{\omega}^{2}x\Bigr]\frac{dx}{\sqrt{x}},
$$
where $\bar{\omega}=\omega/\varepsilon^{1/3}_0$ is the dimensionless frequency.
The coefficient $\mathcal{J}(\bar{\omega})$, which is a function of frequency,
has the sense of the probability density of states.

Finally, using the von Neumann mean ergodic theorem \cite{Neu} and also the Birkhoff
pointwise ergodic theorem \cite{Birk}, we can calculate the explicit  form of the
function $\varrho(\bar{\omega})$:
 \begin{equation}
\varrho(\bar{\omega})=\langle\zeta(t)\rangle=\int_{-\infty}^{+\infty}
\mathcal{P}(\bar{\zeta};\bar{\omega})\bar{\zeta} d\bar{\zeta}=
 \sqrt{\pi}\,\breve{\mathcal{J}}(\bar{\omega})\int_0^{\infty}\sqrt{x}
 \exp\Bigl[-\frac{x^3}{6}-2\,\bar{\omega}^{2}x\Bigr]dx,
\label{2.32b}
\end{equation}
where $\breve{\mathcal{J}}(\bar{\omega})=\mathcal{J}(\bar{\omega})/\varepsilon^{1/3}_0.$\\
Note that the function $\varrho(\bar{\omega})$ has the dimension of frequency.
Following the standard procedures (see in detail \cite{Ashg1}), we can construct a
measure of the functional space $R^\sigma_{\{f\}}$ and, accordingly, to calculate
the functional integral entering into the expression (\ref{2.07c}):
\begin{equation}
\label{2.w3a}
I_1(t)=\biggl\langle \exp\biggl\{\int^t_{-\infty}
\zeta_\sigma(t')dt'\biggr\}\biggr\rangle_{R^\sigma_{\{f\}}}=
\int_{-\infty}^{+\infty}\mathcal{P}^1(\zeta_\sigma,t)d\zeta_\sigma,
\end{equation}
where $\mathcal{P}^1(\zeta_\sigma,t)$ is a function satisfying the following
second-order partial differential equation:
\begin{equation}
\frac{\partial\mathcal{P}^1}{\partial t}=\biggl\{3\zeta+
\bigl(\zeta^2+\omega^2\bigl)\frac{\partial}{\partial \zeta}
+\frac{\varepsilon_0}{2}\frac{\partial^2}{\partial\zeta^2}\biggr\}\mathcal{P}^1.
\label{2.3wv}
\end{equation}

Now it is important to show that the integral (\ref{2.w3a}) converges.
As proven (see Appendix A), in the limit of statistical equilibrium;
$\lim_{t\to\infty}I_1(t)\leq M =const $, or, which is the same thing,
the integral (\ref{2.w3a}) converges. The latter means that the function
$\mathcal{P}^1(\zeta_\sigma,t)$ can be given the meaning of the probability
density and normalized it on unity.

Thus, we have proved that on the scale of the relaxation time $\tau_0$, the
system goes to a statistical equilibrium state and describing by the stationary wave
function (\ref{2.07c}). Obviously, in this case the parameter $\varrho(\bar{\omega})$
is a regular function of the frequency.

\subsection{T\lowercase{he wave function of a massless particle with spin $1$}}
Since the equations in the system (\ref{2.0k2b}) are independent, we can investigate
them separately. For definiteness, consider the first equation of the system
(\ref{2.0k2b}), which describes the $x$ component of QVF.

Representing the wave function in the form:
\begin{equation}
\phi_x^+(\textbf{r})=\phi_x^{+(r)}(\textbf{r})+i\phi_x^{+(i)}(\textbf{r}),
\label{3.020}
\end{equation}
from the first equation of the system (\ref{2.0k2b}), we can get the following two equations:
\begin{eqnarray}
\biggl\{\triangle-\Bigl[\Bigl(\frac{\omega}{c}\Bigr)^2+\frac{\lambda}{r}\Bigr]\biggr\}
\phi_x^{+(r)}(\textbf{r})-\lambda\frac{z-y}{r^2}\phi_x^{+(i)}(\textbf{r})
=0,\,\,\,\,\,
\nonumber\\
\biggl\{\triangle-\Bigl[\Bigl(\frac{\omega}{c}\Bigr)^2+\frac{\lambda}{r}\Bigr]\biggr\}
\phi_x^{+(i)}(\textbf{r})+\lambda\frac{z-y}{r^2}\phi_x^{+(r)}(\textbf{r})=0,\,\,\,\,\,
\label{3.02a}
\end{eqnarray}
where the parameter:
\begin{equation}
\lambda =-\varrho(\bar{\omega})/c<0,
 \label{2.12}
\end{equation}
has the dimension of the inverse distance.

It is easy to show that the equations (\ref{3.02a}) are invariant with
respect to permutations:
$$\phi_x^{+(r)}(\textbf{r})\mapsto \phi_x^{+(i)}(\textbf{r}),\qquad
\phi_x^{+(i)}(\textbf{r})
\mapsto -\phi_x^{+(r)}(\textbf{r}).$$
From this it follows that the solutions $\phi_x^{+(r)}(\textbf{r})$ and
$\phi_x^{+(i)}(\textbf{r})$ globally are  equivalent and differ only by sign.
In other words, the symmetry properties mentioned above make it possible
to obtain two independent equations of the form:
 \begin{eqnarray}
\biggl\{\triangle+\Bigl[-\Bigl(\frac{\omega}{c}\Bigr)^2+|\lambda|\frac{r-(y-z)}{r^2}
\Bigr]\biggr\}\phi_x^{+(r)}(\textbf{r})=0,
\nonumber\\
\biggl\{ \triangle+\Bigl[-\Bigl(\frac{\omega}{c}\Bigr)^2+|\lambda|\frac{r+(y-z)}{r^2}
\Bigr]\biggr\}\phi_x^{+(i)}(\textbf{r})=0.
\label{3.02b}
\end{eqnarray}

Now let us analyze the possibility of obtaining a discrete set of
solutions for wave functions, which can describe a localized state.
For definiteness, we consider the solution of the equation for the
wave function $\phi_x^{+(r)}(\textbf{r})$ on the plane:
\begin{equation}
\label{3.02ab}
r-y+z=\mu r,
\end{equation}
where $\mu$ is a some parameter. The  changing range of this parameter will be
defined below.

Given the equation (\ref{3.02ab}), the first equation in (\ref{3.02b}) can be written as:
\begin{equation}
\label{3.03k}
\Bigl\{\triangle+\Bigl[-\Bigl(\frac{\omega}{c}\Bigr)^2+\frac{|\lambda|\mu}{r}
\Bigr]\Bigr\}\phi_x^{+(r)}(\textbf{r})=0.
\end{equation}
It is convenient to carry out further investigation of the problem in spherical
coordinates. Rewriting (\ref{3.03k}) in the spherical coordinate
system $(x,y,z)\mapsto(r,\theta,\varphi)$, we obtain:
\begin{equation}
\biggl\{\frac{1}{r^{2}}\Bigl[\frac{\partial}{\partial r}\Bigl(r^2\frac{\partial}{\partial r}\Bigr)
+\frac{1}{\sin^2\theta}\frac{\partial^2}{\partial\varphi^2}+\frac{1}{ \sin\theta}
\frac{\partial}{\partial\theta}\Bigl(\sin\theta
\frac{\partial}{\partial\theta}\Bigr)\Bigr]
+\Bigl[-\Bigl(\frac{\omega}{c}\Bigr)^2+ \frac{|\lambda| \mu(\theta,\varphi)}{r}\Bigr]
\biggr\}\phi_x^{+(r)}=0.
\label{3.03al}
\end{equation}
Representing the wave function in the form:
\begin{equation}
\phi_x^{+(r)}(\textbf{r})=\Lambda(r)Y(\theta,\varphi),
\label{3.03b}
\end{equation}
we can conditionally divide the variables in the equation (\ref{3.03al}) by writing
it in the form:
\begin{eqnarray}
r^2\Lambda ^{''}+2r\Lambda ^{'}+\bigl[-(\omega/c)^2 r^2+|\lambda|\mu(\theta,\varphi)r
-\nu\bigr]\Lambda =0,\quad
\label{3.02bt}
\end{eqnarray}
\vspace {-2mm}
and, respectively;
\begin{eqnarray}
\frac{1}{\sin\theta}\Bigl\{
\frac{1}{\sin\theta}\frac{\partial^2}{\partial\varphi^2}+\frac{\partial}{\partial\theta}
\Bigl(\sin\theta\frac{\partial}{\partial\theta}\Bigr)\Bigr\}Y+\nu Y=0,
\label{3.02bt'}
\end{eqnarray}
where $\Lambda^{'}=d\Lambda/dr$ and $\nu$ is a constant, which can  represented in
the form $\nu=l(l+1)$, in addition, $l=0,1,2...$

Note that the conditional separation of variables means to impose an
additional condition on the function $\mu(\theta,\varphi)=const$.
Writing equation (\ref{3.02ab}) in spherical coordinates, we obtain the
following trigonometric equation:
\begin{equation}
\label{3.02abc}
\mu(\theta,\varphi)=1-\sin\theta\sin\varphi +\cos\theta.
\end{equation}
Analysis of the equation (\ref{3.02abc}) shows that   $\mu\in[(1-\sqrt{2}), (1+\sqrt{2})]$.
The solution of the equation (\ref{3.02bt'}) is well known, these are spherical Laplace
functions $Y_{l,m}(\theta,\varphi)$, where $m=0,\pm1,...,\pm l$.

As for the equation (\ref{3.02bt}), we will solve it for a fixed value
$\mu$, which is equivalent to the plane cut of the three-dimensional solution.
In particular, we will seek a solution $\Lambda(r)$ tending to finite value
for $r\to 0 $ and, respectively, to zero at $r\to\infty$.

For a given  parameter $\mu_0>0$,  we can write the equation (\ref{3.02bt})
 in the form:
\begin{eqnarray}
\frac{d^2\Lambda}{d\rho^2}+\frac{2}{\rho}\frac{d\Lambda}{d\rho}+\Bigl[-\beta^2 +\frac{2}{\rho}-
\frac{l(l+1)}{\rho^2}\Bigr]\Lambda=0,
\label{3.02btz}
\end{eqnarray}
where $\rho=r/a_p$ and $a_p=2/(|\lambda|\mu_0)$ denotes the characteristic
spatial dimension of a hypothetical massless Bose particle with spin projection +1,
in addition, the parameter $\beta$, which further will be play a key role
for finding a discrete set of solutions is determined by the expression
$\beta=(\omega a_p/c)$.

It is important to note that from the symmetry and non-coincidence of the components
$\phi^{+(r)}_x$ and $\phi^{+(i)}_x$, it follows that $\mu_0 = 2$. This fact will be taken
into account in further calculations.

As well-known the solution of the equation (\ref{3.02btz}) describes the radial
wave function of the hydrogen-like system, which is written in the form \cite{Land}:
\begin{eqnarray}
\label{3.03tz}
\Lambda_{nl}(r)=\frac{(b)^{3/2}
(br)^le^{-br/2}}{\sqrt{2n(n-l-1)!(n+l)!}}L_{n-l-1}^{2l+1}(br),\quad
\end{eqnarray}
where $b=(2/na_p)$. In addition, the functions $L_{n}^{k}(x)$ are associated
Laguerre polynomials orthogonal to $[0,\infty)$ with respect to the weight function
$x^ke^{-x}$ and, respectively, satisfy the condition normalization:
$$
\int^\infty_0 x^ke^{-x}L_{n}^{k}(x)L_{m}^{k}(x)dx=\frac{(n+k)!}{n!}\delta_{mk},
$$
where $\delta_{mk}$ is the Kronecker delta (more detail see \cite{Abram}). 

Note that the solution (\ref{3.03tz}) takes place if the following condition is satisfied:
\begin{equation}
\label{3.03}
n_r+l+1=n+l=\beta^{-1},\qquad n_r=0,1,2...,
\end{equation}
where $n_r$ is the radial quantum number, $n$ is the principal quantum number and
$l$ denotes the quantum number of the angular momentum, which is limited $l\leq n-1$.

In other words, the quantization  condition is the integer value of the term
$\beta^{-1} $, which implies satisfying the following conditions:
 \begin{eqnarray}
 \label{3.04ak}
 \bigl[\beta^{-1}\bigr]=\Bigl[\frac{\breve{\varrho}(\bar{\omega}) }{\bar{\omega}}\Bigr]=n,\qquad
 \bigl\{\beta^{-1}\bigr\}=\Bigl\{ \frac{\breve{\varrho}(\bar{\omega})}{\bar{\omega}}\Bigr\}=0,
 \end{eqnarray}
where $\breve{\varrho}(\bar{\omega})=\varepsilon^{1/3}_0\varrho(\bar{\omega})$ is a
dimensionless function, the brackets $[...]$ and $\{...\}$ denote the integer and
fractional parts of the function, respectively.

As follows from the calculations (see Fig. 1),
in the frequency range $\bar{\omega}\in\{0.05,0.34\}$ there are 8 points, that are highlighted
in red, satisfy the quantization conditions (\ref{3.04ak}). The latter means that in
specified frequency range there are only eight quantum states, however the number of these
states is growing at $\bar{\omega}\to 0.$

\textbf{Table 1}. \emph{The average-statistical dimensionless frequency of the system
in different quantum states  (see condition (\ref{3.03})).}
$$
\begin{array}{cccccccc}
\beta^{-1}=n+l\,= 1 & 2 & 3 & 4 & 5 & 6& 7 & 8\,... \\
\qquad 10^{-2}\times\bar{\omega}= 34 &20 &14 & 10 & 9 &7 & 6 &5\,...
\end{array}
 $$
\begin{figure}
\includegraphics[width=110mm]{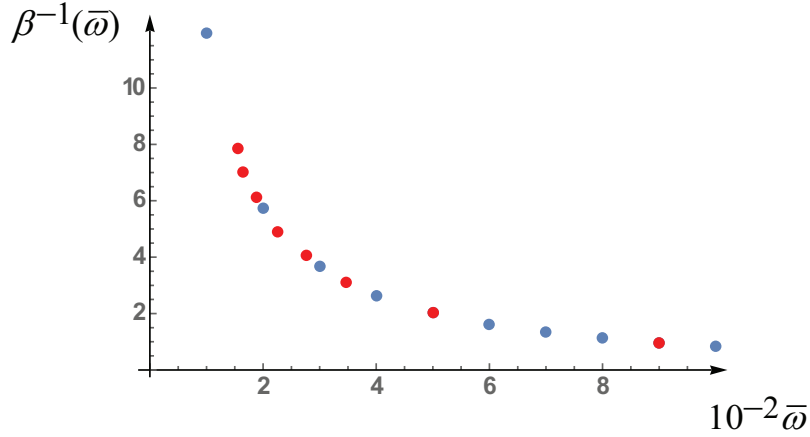}
\caption{\emph{The dependence of a quantity $\beta^{-1}(\bar{\omega})$ on the
dimensionless frequency. As calculations show, in the frequency range under
consideration (see table) there are only eight values of $\beta^{-1}$ (red points),
for which the quantization conditions (\ref{3.03})-(\ref{3.04ak}) are satisfied.
The blue dots denote such states for which the quantization conditions are not satisfied.}}
\label{Fig.1}
\end{figure}
Now we consider the problem of localization of the solution $\phi^{+(r)}_x$. Taking into
account the fact that $\mu_0=2$, the equation (\ref{3.02abc}) can be written in the form:
\begin{equation}
\label{3.03by}
1+\sin\theta\sin\varphi -\cos\theta=0.
\end{equation}
In particular, as follows from the equation (\ref{3.03by}), all solutions (\ref{3.03tz})
are localized on the manifold  $S^r_x(\theta,\varphi)\cong(-Y,Z)$ (one-quarter of the
plane $\{Y,Z\}$, where the bracket $\{.\,,.\}$ denotes the entire plane (see Fig. 2).

The imaginary part of the wave function $\phi^{+(i)}_x$ is calculated similarly
and has the same form, but in this case the solution must satisfy the following
trigonometric equation:
\begin{equation}
\label{3.03ba}
1-\sin\theta\sin\varphi+\cos\theta=0.
\end{equation}
Obviously, the equation (\ref{3.03ba}) defines another quarter of the plane
$S^i_x(\theta,\varphi)\cong(Y,-Z)\subset\{Y,Z\}$, on which the solution
$\phi^{+(i)}_x$ is localized. As for the wave function
 $\phi^{+}_x\bigl(\phi^{+(i)}_x,\phi^{+(r)}_x\bigr)$,
then it is localized on the manifold $S^r_x\cup S^i_x\subset\{Y,Z\}$.

A similar investigation for  the projections of the wave function
$\bigl\{\phi^{+(r)}_y,\phi^{+(i)}_y\bigr\}$ and $\bigl\{\phi^{+(r)}_z,\phi^{+(i)}_z\bigr\}$
shows that the separation of variables in corresponding equations is possible taking into
account the following algebraic equations:
\begin{eqnarray}
\label{3.07k}
\bigl\{\phi^{+(r)}_y,\phi^{+(i)}_y\bigr\}:\quad 1\pm\cos\theta\sin\varphi\mp\cos\theta=0,\qquad
\nonumber\\
\bigl\{\phi^{+(r)}_z,\phi^{+(i)}_z\bigr\}:\quad 1\pm\sin\theta\sin\varphi\mp\cos\theta\sin\varphi=0.
\end{eqnarray}
Analysis of the equations (\ref{3.07k}) shows that the projections of the wave
function $\bm\phi^{+}(\textbf{r}) $ are localized on the following manifolds;
$\bigl\{\phi^{+(r)}_y[S^{r}_y\cong(-X,Z)],
\,\phi^{+(i)}_y[S^{i}_y\cong(X,-Z)\bigr]\bigr\}$ and $\bigl\{\phi^{+(r)}_z[S^{r}_z\cong(-X,Y)],\,
\phi^{+(i)}_z[S^{i}_z\cong(X,-Y)]\bigr\}$.
\begin{figure}
\includegraphics[width=90mm]{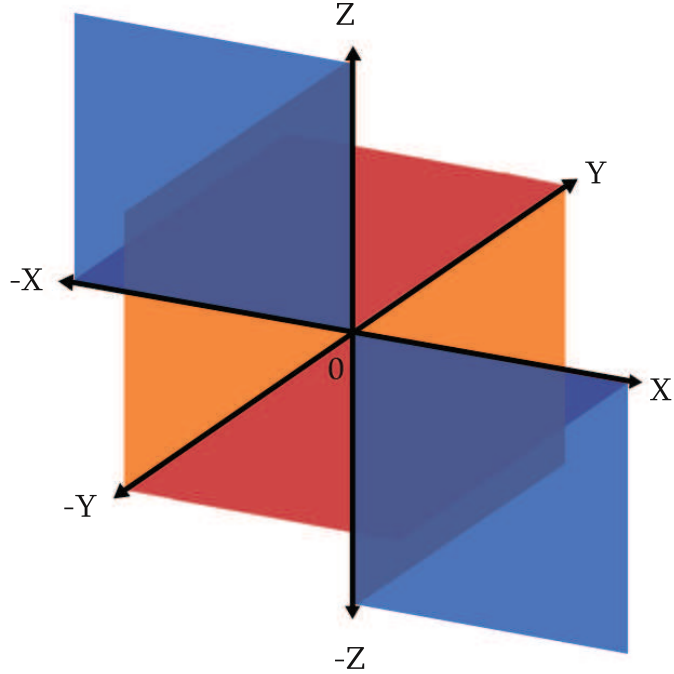}
\caption{\label{fig:epsart} \emph{The coordinate system $\{X,Y,Z\}$ divides
the three-dimensional space into eight spatial regions using three planes. The
boson of a vector field (hion) with projection of spin +1 is a two-dimensional
structure consisting of six components localized on the following manifolds
$\phi^+_x[(-Y,Z)\cup(Y,-Z)],$ \,$\phi^+_y[(-X,Z)\cup(X,-Z)]$ and
$\phi^+_z[(-X,Y)\cup(X,-Y)]$, respectively.}}
\end{figure}
\\
\emph{\textbf{The theorem is proved.}}
\\
Thus, we have proved the possibility of the formation of a stable massless Bose
particle with a spin of 1 as a result of random fluctuations of the QVF.
As can be seen the obtained solutions (\ref{3.03tz}) combine the properties of quantum
mechanics and the theory of relativity and, respectively, maximally reflect to the ideas
of \emph{string theory}. It is interesting to note that the \emph{ground state} of
the vector boson characterized the highest frequency. In the future we will call
the particle of a vector field \emph{hion}.

\section{Q\lowercase{uantum distribution in different \emph{hion} states}}

Let us consider the solution of the equation (\ref{3.02btz}) in the \emph{ground state}.

Taking into account (\ref{3.03b}) and (\ref{3.03tz}), we can get the following
solution: 
$$\phi^{+(r)}_x(\textbf{r})=
\phi^{+(r)}_{x(1,0,0)} (\textbf{r})=\Lambda_{10}(r)Y_{0,0}(\theta,\varphi)
$$
where
\begin{equation}
\label{3.03a}
\Lambda_{10}(r)=Ca_p^{-3/2}e^{-r/a_p},\qquad
 Y_{0,0}(\theta,\varphi)=\frac{1}{2\sqrt{\pi}},
\end{equation}
in addition, the indices $(1,0,0)$ of the wave function denote   the  quantum numbers $(n,l,m)$,
accordingly, the constant $C$ is defined below from the normalization
condition of the wave function, in addition, $a_p$ is the characteristic spatial
dimension of the vector boson in the \emph{ground state}, which can be calculated
taking into account the equations (\ref{2.32b}) and (\ref{2.12}):
\begin{equation}
a_p= |\lambda|^{-1}=c/\varrho(\bar{\omega})= 2^{1/3} c\,\omega^{-1}.
\label{3.05at}
\end{equation}
Recall that in (\ref{3.05at}) the frequency $\omega=\varepsilon^{1/3}_0\bar{\omega}$,
where dimensionless frequency of the \emph{ground state} is equal $\bar{\omega}=0.34$
(see table and Fig. 1). In the framework of the developed approach, it is impossible
to determine the constant $a_p$, since the speed $c$ and fluctuations power
$\varepsilon_0$ remain free parameters of the theory. Apparently, these
parameters will have to be refined experimentally and introduced
into the theory as fundamental constants.

As for the wave function $\phi^{+(i)}_{x(1,0,0)}(\textbf{r})$, it is also described by the
expressions (\ref{3.03a}), but with the only difference that in this case the wave function
is localized on the manifold $S^i_x$. In a similar way one can obtain solutions for the
wave functions $\phi^+_{y(1,0,0)}$ and $\phi^+_{z(1,0,0)} $ localized on the corresponding
manifold.

Now we can write down the normalization condition for the full wave function:
\begin{equation}
\label{3.04b}
  \int\bm\phi^+\bigl(\bar{\bm\phi}^+\bigr)^TdV=1, \qquad dV=dxdydz,
\end{equation}
where $\bigl(\bar{\bm\phi}^+\bigr)^T=
\bigl(\bar{\bm\phi}^+_x,\bar{\bm\phi}^+_y,\bar{\bm\phi}^+_z\bigr)$.

Considering that the projections of the full wave function
$\bm\phi^{+}(\textbf{r})$ are localized on different non-intersecting manifolds
and the definition (\ref{3.04b}), we can write:
$$
 \int\phi^+_x\bar{\phi}^+_xdV=\int\phi^+_y\bar{\phi}^+_ydV=\int \phi^+_z\bar{\phi}^+_zdV= {1}/{3}.
$$

Below as an example, we  will calculate the first term of the integral, considering the case of the
\emph{ground state}. Taking into account that the wave function $\phi^{+}_x$ can be
represented in the form $\phi^{+}_x=\phi^{+(r)}_x+i\phi^{+(i)}_x$, we can write:
\begin{eqnarray}
\label{3.05}
\int\phi^+_x\bar{\phi}^+_xdV=\int\bigl(|\phi^{+(r)}_x|^2+|\phi^{+(i)}_x|^2\bigr)dV
=2\int|\phi^{+(r)}_x|^2dV=
\nonumber\\
2\int|\phi^{+(i)}_x|^2dV=\frac{a_p}{2\pi}\int_{(-Y,Z)}|
\Lambda_{10}(\rho)|^2 dydz=\frac{1}{3},
\end{eqnarray}
where $\varrho=r(0,y,z)= \sqrt{y^2+z^2}$ denotes radius-vector $r$ on the plane $S^r_x=(-Y,Z)$,
in addition, in calculating the integral, we assume that the wave function in the
direction $x$ perpendicular to the plane $S^r_x$ is the Dirac delta function.

Taking into account (\ref{3.03a}), we can calculate the integral in the expression  (\ref{3.05}):
\begin{equation}
\label{3.05a}
\int_{(-Y, Z)}|\Lambda_{10}(\rho)|^2dydz=\frac{\pi}{8a_p}.
\end{equation}
Considering (\ref{3.05}) and (\ref{3.05a}), we can determine the normalization constant
of the wave function (\ref{3.03a}), which is equal to $C=4/\sqrt{3}$. Note that in
a similar way one can obtain the \emph{hion} wave function with the spin projection
-1 (see Appendix B).

Now we can calculate the probability distribution of the \emph{hion's} $x$-projection
in the \emph{ground state}. Using (\ref{3.03a}), we obtain the following
expression for the probability distribution on the surface element $dS$:
\begin{equation}
 W(\rho)dS=\frac{1}{4\pi}{C'}^2e^{-2\rho}{\rho}d{\rho}d\vartheta,
\label{3.06t}
\end{equation}
where $C'=C/a_p^{3/2}$ and $dS =\rho d\rho d\vartheta$.

Recall that the angle $\vartheta$ coincides with the angle $\theta$ on the fixed plane.
\begin{figure}
\includegraphics[width=100mm]{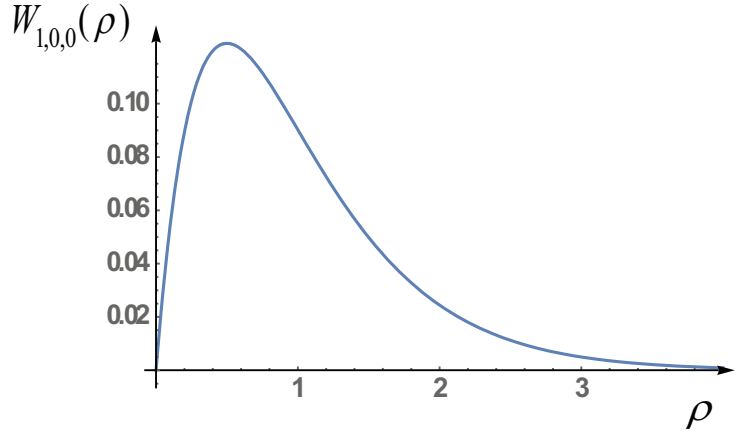}
\caption{\emph{The probability distribution of hion in the
\emph{ground state} depending on the radius. The distance $\rho_0=1/2$, or more
precisely $\varrho_0=a_p/2$, at which the maximum value of the amplitude of the
hion probability is reached.}}
\end{figure}
Integrating the expression (\ref{3.06t}) by the angle $\vartheta\in[0,\pi/2]$, we
obtain the probability distribution of the \emph{ground state} depending on radius:
\begin{equation}
W_{1,0,0}(\rho)=\frac{2}{3}\rho e^{-2\rho}.
\label{3.a06t}
\end{equation}
Finally, calculating the expression (\ref{3.a06t}), we find that for the value $\rho_0=1/2$
and, respectively, for $r(0,x,y) =\rho=a_p/2$, the probability distribution has a
maximum (see Fig. 3).

Now we consider the first three excited quantum states, which are characterized by the principal
quantum number $n=2$. Using the solution (\ref{3.03tz}), we can write the explicit
form of these wave functions:
 \begin{eqnarray}
\phi^{+(r)}_{x(2,0,0)}\,\,=\,\frac{1}{3\sqrt{2}}\frac{2-\rho}{a_p^{3/2}}e^{-\rho/2}Y_{0,0},
  \nonumber\\
 \phi^{+(r)}_{x(2,1,0)}\,\,=\,\frac{2}{3\sqrt{3}}\frac{\rho}{a_p^{3/2}}\,\,e^{-\rho/2}Y_{1,0},
 \nonumber\\
\phi^{+(r)}_{x(2,1,\pm1)}=\frac{1}{3\sqrt{3}}\frac{\rho}{a_p^{3/2}}e^{-\rho/2}Y_{1,\pm1},
 \label{3.07}
\end{eqnarray}
where
$$
Y_{0,0}=\frac{1}{2\sqrt{\pi}},\qquad Y_{1,0}=\sqrt{\frac{3}{4\pi}}\cos\vartheta,
\qquad
Y_{1,\pm1}=\mp e^{\pm i\varphi}\sqrt{\frac{3}{8\pi}}\sin\vartheta.
$$
\begin{figure}
\center\includegraphics[width=100mm]{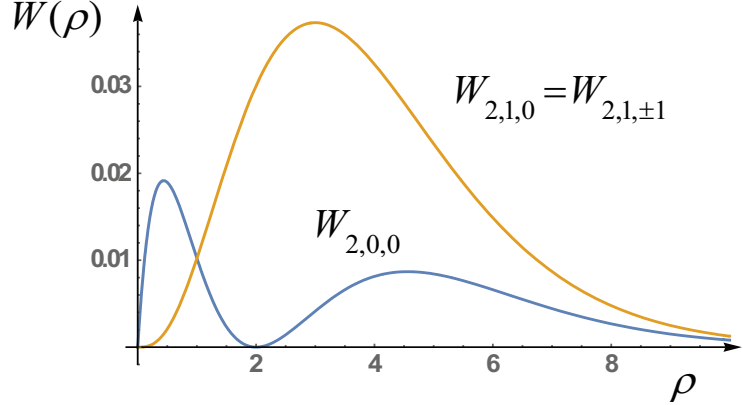}
\caption{\emph{The probability distributions of the first four excited states of
the \emph{hion} depending on radius. The orange curve in the graph shows the
probability distributions in various three excited states.} }
\end{figure}
Taking into account expressions (\ref{3.07}), we can construct a radial probability distribution
for the first four excited states of the \emph{hion}:
$$
W_{2,0,0}=\frac{1}{6^2}(2-\rho)^2\rho e^{-\rho}, \qquad
W_{2,1,0}=W_{2,1,\pm1}=\frac{1}{6^2}\rho^3e^{-\rho}.
$$
Recall that at deriving of expressions for the probability distributions $W_{2,1,0}(\rho)$,
$W_{2,1,+1}(\rho)$ and $W_{2,1,-1}(\rho)$, averaging over the angle $\vartheta$ is performed.
Note that the probability distributions  (see FIG. 4), and also  the energies of considered
three states coincide. In particular, the quantum state described by the wave function
$\phi^{+(r)}_{x(2,0,0)}$ has the energy $\mathcal{E}_{2,0,0}=-0.2\hbar \varepsilon^{1/3}_0$,
whereas  three different quantum states $\phi^{+(r)}_{x(2,1,0)}$, $\phi^{+(r)}_{x(2,1,+1)}$ and
 $\phi^{+(r)}_{x(2,1,-1)}$ are characterized by the same energy $\mathcal{E}_{2,1,0}=
 \mathcal{E}_{2,1,\pm1}=-0.14\hbar \varepsilon^{1/3}_0$.

\section{T\lowercase{he state of the \emph{hion} on the next scale of relaxation }}
So far we have investigated the possibility of the formation of \emph{hion}
as a result of the continuously fluctuations of QVF on the first phase of relaxation,
while such phases can be a few. In this connection, the natural question arises:
namely, how  the state of the \emph{hion} changes if we take into account random
fluctuations of QVF of the next order, ie, consider the change of the particle on
the next evolutionary scale $ \{\tau_1,{\bm\varepsilon}_1\}$.

Let us consider the evolution of \emph{hion} with the spin projection +1 taking
into account the influence of the random environment in the framework of SDE of the type:
\begin{equation}
\partial_{t} {\breve{\bm{\psi}}^{+}}(\textbf{r},t)-c\bigl(\textbf{S}\cdot \bm\nabla\bigr)
\bm{\breve{{\psi}}^{+}}(\textbf{r},t)=\bm\eta^+(s),
\label{4.s0}
\end{equation}
 and also the equation:
\begin{equation}
\nabla\bm{\breve{{\psi}}^{+}}(\textbf{r},t)=0,\qquad t\in(-\infty,+\infty),
\label{4.s1}
\end{equation}
where $\bm\eta^+(s)=(\eta_x^+,\eta_y^+,\eta_z^+)$ denotes the generator of random forces,
and $ds^2=c^2dt^2-dx^2-dy^2-dz^2$ is the 4$D$-interval in which these random influences are carried.
The equation (\ref{4.s0}) can be represented in matrix form:
\begin{equation}
\left[
\begin{array}{ccc}
ict & -z & y\\
z& ict & -x \\
-y& x & ict \\
\end{array}
\right]\cdot
\left[
  \begin{array}{ccc}
    \dot{\breve{{\psi}}}_x^+\\
   \dot{\breve{{\psi}}}_y^+ \\
  \dot{\breve{{\psi}}}_z^+\\
  \end{array}
\right]=s\left[
  \begin{array}{ccc}
  \eta_x^+ \\
  \eta_y^+ \\
  \eta_z^+\\
  \end{array}
\right],\quad
\label{4.s2}
\end{equation}
and the equation (\ref{4.s1}), respectively, in the form:
\begin{equation}
  x \dot{\breve{{\psi}}}^{+}_x+ y \dot{\breve{{\psi}}}^{+}_y+ z\dot{\breve{{\psi}}}^{+}_z=0,
  \qquad \dot{\breve{{\psi}}}^{+}_\sigma=\partial{\breve{{\psi}}}^{+}_\sigma/\partial{s}.
\label{4.s03}
\end{equation}

For further constructions, the system of equations (\ref{4.s2})-(\ref{4.s03}) must
be reduced to the canonical form:
\begin{equation}
\dot{\breve{{\psi}}}^{+}_\sigma(\bar{s};\textbf{r},t)=\bigl\{b^+_\sigma(\textbf{r},t)+
d^+_\sigma(\textbf{r},t)\bigr\}\bar{s}^{-1}\eta(\bar{s}),
\label{4.s04}
\end{equation}
where $\bar{s}=s/a_p$ and  $\eta(\bar{s})=\bar{s}^{-1}\eta_x=\bar{s}^{-1}\eta_y=\bar{s}^{-1}\eta_z$,
in addition:
\begin{eqnarray}
b^+_x=\frac{z-y}{a_p},\qquad d_x^+=\frac{c^2t^2-x^2-xy-xz}{a_pct},
\nonumber\\
b^+_y=\frac{x-z}{a_p},\qquad d^+_y=\frac{c^2t^2-y^2-xy-yz}{a_pct},
\nonumber\\
b^+_z=\frac{y-x}{a_p},\qquad d^+_z=\frac{c^2t^2-z^2-xz-yz}{a_pct}.
\label{4.s05}
\end{eqnarray}
For simplicity, we will use a random generator that satisfies the conditions of \emph{white noise}:
\begin{equation}
\langle\eta(\bar{s})\rangle=0,\qquad \langle\eta(\bar{s})\eta(\bar{s}')\rangle=
2\varepsilon_1\delta(\bar{s}-\bar{s}'),
\label{4.s06}
\end{equation}
where $\varepsilon_1=\varepsilon^r_1+i\varepsilon^i_1$ and $\varepsilon^r_1=\varepsilon^i_1,$
in addition, it is assumed that the bracket $\langle...\rangle$ means averaging over the
relaxation time $\tau_1$.

The joint probability distribution of QVF can be represented in the form
(see \cite{AshG}):
\begin{eqnarray}
\mathcal{P}(\{\breve{\bm\psi}^+\},\bar{s};\textbf{r},t)=\prod_{\sigma}
\bigl\langle\delta\bigl(\breve{\bm\psi}_\sigma^+(\bar{s};\textbf{r},t)-
{\bm\breve{\phi}}_{\sigma}^+\bigr)\bigr\rangle,
\label{4.s07}
\end{eqnarray}
where the set of wave functions  $\{\breve{\bm\psi}^+\}=({\breve{\psi}}^+_x,
{\breve{\psi}}^+_y,{\breve{\psi}}^+_z)$ denotes vacuum fields, $\{{\bm\breve{\phi}}^+\}=
({\breve{\phi}}^+_x,{\breve{\phi}}^+_y,{\breve{\phi}}^+_z)$ and
 ${\breve{\phi}}_{\sigma}^+(\textbf{r})={\breve{\psi}}_{\sigma}^+(\bar{s};\textbf{r},t)|_{s\to0}$.
In addition, in (\ref{4.s07}) the function $\delta(\breve{\bm\psi}_\sigma^+(\bar{s};\textbf{r},t)-
\breve{\bm{\phi}}_{\sigma}^+)$ denotes the Dirac delta function in the three-dimensional
Hilbert space, in addition, by default we will assume that the wave function is dimensionless,
i.e. it is multiplied by a constant value
$a_p^{3/2}$  (see (\ref{3.03a})).

Now using the system of SDE (\ref{4.s04}) and (\ref{4.s05})-(\ref{4.s06}), for the
conditional probability (\ref{4.s07}) the following second order partial differential
equation can be obtained \cite{Ashg1}:
\begin{eqnarray}
\biggl\{\frac{\partial }{\partial\bar{s}}-\frac{1}{2} \sum_{\sigma}
\varepsilon_{1\sigma}^+(\textbf{r},t) \frac{\partial^2 }{\partial{\breve{\psi}}_\sigma^+
\partial\bar{\breve{\psi}}_\sigma^+}\biggr\}\mathcal{P}=0,
\qquad
\frac{\partial^2 }{\partial{\breve{\psi}}_\sigma^+\partial\bar{\breve{\psi}}_\sigma^+}
=\frac{\partial^2 }{\partial\bigl[{\breve{\psi}}_\sigma^{+(r)}\bigr]^2}
+\frac{\partial^2 }{\partial\bigl[{\breve{\psi}}_\sigma^{+(i)}\bigr]^2},
\label{4.s08}
\end{eqnarray}
where $\bar{\breve{\psi}}_\sigma^+$ denotes the complex conjugate of the wave function
$\breve{\psi}_\sigma^+$ and $\varepsilon_{1\sigma}^+(\textbf{r},t)=
\varepsilon_1[b_\sigma^+(\textbf{r},t)+
d_\sigma^+(\textbf{r},t)]^{2}$, which is a dimensionless quantity denoting the fluctuations
power. In the equation (\ref{4.s08}) the following
notations also  are made; $\breve{\psi}_\sigma^{+(r)}=\emph{Re}(\breve{\psi}_\sigma^+)$ and
$\breve{\psi}_\sigma^{+(i)}=\emph{Im}(\breve{\psi}_\sigma^+)$.

The general solution of the equation (\ref{4.s08}) is convenient to represent in
the integral form:
\begin{widetext}
\begin{eqnarray}
\mathcal{P}(\{\breve{\bm\psi}^+\},\bar{s}\,;\textbf{r},t)=\int_{\Xi^3}
R(\{\breve{\bm\phi}^+\})\prod_{\sigma}
\exp\biggl\{-\frac{(\breve{\phi}_{\sigma}^+-{\breve{\psi}}_\sigma^+)(\breve{\bar{\phi}}_{\sigma}^+
-\bar{{\breve{\psi}}}_\sigma^+)}{2\bar{s}\varepsilon_{1\sigma}^+}\biggr\}
\frac{d\breve{\phi}_{\sigma}^+}{\sqrt{2\pi\bar{s}\varepsilon_{1\sigma}^+}},
\label{4.s09}
\end{eqnarray}
\end{widetext}
where $d\breve{\phi}_{\sigma}^+=d\breve{\phi}_{\sigma}^{+(r)}
d\breve{\phi}_{\sigma}^{+(i)}$ and the function $R(\{\breve{\bm\phi}^+\})$ denotes the initial condition of
the equation (\ref{4.s08}) at $s=0$, before switching on the interaction with the
random environment. Since before switching on the interaction, the \emph{hion}
(the vector-boson) is in a pure quantum state, ie, in the Hilbert space is determined
by a fixed vector  $\bm\phi^+$,  then we can put; $R(\{\breve{\bm\phi}^+\})=
\mathcal{{P}}^0(\{{\bm\phi}^+\})$, where $\mathcal{{P}}^0(\{{\bm\phi}^+\})$ has
the sense of the distribution \emph{hion}, which is defined as follows:
\begin{equation}
 \mathcal{{P}}^0(\{\bm\phi^+\})=||\bm{\phi}^+||^2=\left[
  \begin{array}{ccc}
\mathcal{{P}}^0_x(\phi^+_x) \\
\mathcal{{P}}^0_y(\phi^+_y)\\
\mathcal{{P}}^0_z(\phi^+_z)\\
  \end{array}
\right],
\label{4.s11}
\end{equation}
where
\begin{equation}
\mathcal{{P}}^0_\sigma(\phi^+_\sigma)=\frac{1}{2}||\phi^+_\sigma||^2=
\frac{1}{2}\sum_{\varpi=r,i}||\phi^{+(\varpi)}_\sigma||^2,\qquad
||\phi^{+(\varpi)}_\sigma||^2=\phi^{+(\varpi)}_\sigma
\bar{\phi}^{+(\varpi)}_\sigma.
\label{4.s12}
\end{equation}

Substituting the expressions  (\ref{4.s11})-(\ref{4.s12}) into (\ref{4.s09}) and
integrating over the variables $ \bm{\breve{\phi}}^+$ within
$[{\breve{\psi}}^{+(\varpi)}_\sigma,\phi^{+(\varpi)}_\sigma]$,
we obtain the expression for the deformation of the initial quantum distribution
$\mathcal{{P}}^0(\{{\bm\phi}^+\})$, taking into account the evolution of \emph{hion}
in a random environment:
\begin{equation}
\mathcal{P}(\{\bm\phi^+\},\{\breve{\bm\psi}^+\},\bar{s}\,;\textbf{r},t)
=\left[\begin{array}{ccc}
\mathcal{{P}}_x(\phi^+_x,\breve{\psi}^+_x;\bar{s},t) \\
\mathcal{{P}}_y(\phi^+_y,\breve{\psi}^+_y;\bar{s},t)\\
\mathcal{{P}}_z(\phi^+_z,\breve{\psi}^+_z;\bar{s},t)\\
  \end{array}
\right],
\label{4.s13}
\end{equation}
 where
\begin{eqnarray}
\mathcal{{P}}_\sigma(\phi^+_\sigma,\breve{\psi}^+_\sigma;\bar{s},t)=\frac{1}{2}
\sum_{\varpi=r,i}||\phi^{+(\varpi)}_\sigma||^2
F_\sigma^{+(\varpi)}.
\label{4.s14}
\end{eqnarray}
Note that the function $F_\sigma^{+(\varpi)}$ characterizes the deformation
of the initial distribution:
$$
F_\sigma^{+(\varpi)}\bigl(\phi^{+(\varpi)}_\sigma,
\breve{\psi}^{+(\varpi)}\bigr)=\frac{1}{2}
\Biggl\{1+\mathrm{erf}\biggl[\frac{\phi^{+(\varpi)}_\sigma-
\breve{\psi}^{+(\varpi)}_\sigma}{\sqrt{2\bar{s}\varepsilon_{1\sigma}^+}}\biggr]\Biggr\}.
$$
Integrating  (\ref{4.s09}) taking into account  (\ref{4.s13})-(\ref{4.s14}), we obtain
the quantum distribution of \emph{hion} with consideration of the
random influence of an environment. It is easy to see that before the relaxation,
the 4$D$-interval is zero, i.e. $s=0 $ and, accordingly, the deformation coefficient
$F=1$, as expected.

By similar reasoning, we can calculate the deformation of the
\emph{hion} state vector:
\begin{eqnarray}
\mathcal{D} \phi^{+}_\sigma=
 \bigl\{\phi^{+(r)}_\sigma F_\sigma^{+(r)}+i\phi^{+(i)}_\sigma F_\sigma^{+(i)}\bigr\}.
\label{4.s15}
\end{eqnarray}

Thus, it is obvious that the deformation of the quantum state \emph{hion} leads
to a breaking of symmetry, which  leads to spontaneous transitions from \emph{ground state}
to another, massless, and also mass states.

It is important to note that despite the fact that \emph{hion} is deformed
under the influence of random influences of the environment, nevertheless the full
probability is conserved. In particular, if to integrate the representation (\ref{4.s09})
over the fields $\{\breve{\bm\psi}^+\}$, then, obviously, we can get:
$$
\int_{\Xi^3}\mathcal{P}(\{\breve{\bm\psi}^+\},\bar{s}\,;\textbf{r},t)d\{{\breve{\bm\psi}^+}\}=
 \int_{\Xi^3} \mathcal{P}^0(\{\bm\phi\}) d\{{\bm\phi}\}=
\left[\begin{array}{ccc}
1/3\\
1/3\\
1/3\\
\end{array}
\right],
$$
where $d\{{\breve{\bm\psi}^+}\}=d\breve{\psi}^+_xd\breve{\psi}^+_y d\breve{\psi}^+_z$.

Recall that integration over the Hilbert space $\Xi^3$ is equivalent
to integration over the configuration space  $\mathbb{R}^3$.
This is an obvious proof  that the probability is preserved.

\section{F\lowercase{ormation of singlet and triplet pairs of \emph{hions} }}
At the second stage of relaxation in the ensemble \emph{hions}, the formation
of singlet and triplet states are possible by entangling them \cite{Einst}.

As it is well known \cite{Bell}, there are four possible entangled states
of the so-called Bell states, which can be represented as:
\begin{eqnarray}
\bm\phi^{\updownarrow}_\mp(\textbf{r}_+,\textbf{r}_-)= \frac{1}{\sqrt{2}}
\bigl\{|\uparrow\rangle_1\otimes|\downarrow\rangle_2\mp|\downarrow\rangle_1
\otimes|\uparrow\rangle_2\bigr\},
\nonumber\\
\bm\phi^{\upuparrows}_\mp(\textbf{r}_+,\textbf{r}_-)= \frac{1}{\sqrt{2}}
\bigl\{|\uparrow\rangle_1\otimes|\uparrow\rangle_2\mp|\downarrow\rangle_1
\otimes|\downarrow\rangle_2\bigr\},
 \label{4.03b}
\end{eqnarray}
where the radius vectors $\textbf{r}_+$ and $\textbf{r}_-$ determine positions
of the first and second \emph{hions}, respectively. Note that the first
equation denotes possible two singlet states, and the second - two triplet
states. In (\ref{4.03b}) also the following notations are made:
\begin{eqnarray}
|\uparrow\rangle_1=\bm\phi^+(\textbf{r}_+),\qquad |\downarrow\rangle_2=
\bigl[\bm\phi^-(\textbf{r}_-)\bigr]^T,
\nonumber\\
|\downarrow \rangle_1=\bar{\bm\phi}^+(\textbf{r}_+), \qquad |\uparrow\rangle_2=
\bigl[\bar{\bm\phi}^-(\textbf{r}_-)\bigr]^T,
 \label{4.z03b}
\end{eqnarray}
where we recall that the wave functions $|\uparrow\rangle_1$ and $|\uparrow\rangle_2$ denote
the pure states of \emph{hions} with the spin projections +1 and -1, respectively.
In (\ref{4.03b}) the dash $^-$ over a wave function denotes complex conjugation, $ [...]^T $
denotes the transposed vector and the symbol  $\otimes$, respectively, denotes the tensor
product between the vectors.

The explicit form of a direct tensor product between vectors with opposite
spins have the following form:
$$
\textbf{A}^\updownarrow=|\uparrow\rangle_1\otimes| \downarrow\rangle_2 =
\left[
\begin{array}{ccc}
{\phi}_x^+\\
{\phi}_y^+\\
{\phi}_z^+\\
  \end{array}
\right]
\left[
\begin{array}{ccc}
 {\phi}_x^-\,\,
 {\phi}_y^- \,\,
 {\phi}_z^-
  \end{array}
\right]=\left[
  \begin{array}{ccc}
{\phi}_x^+{\phi}_x^- \,\,\, {\phi}_x^+{\phi}_y^-\,\,\,{\phi}_x^+{\phi}_z^-\\
{\phi}_y^+{\phi}_x^- \,\,\, {\phi}_y^+{\phi}_y^-\,\,\,{\phi}_y^+{\phi}_z^- \\
{\phi}_z^+{\phi}_x^- \,\,\, {\phi}_z^+{\phi}_y^-\,\,\,{\phi}_z^+{\phi}_z^-\\
  \end{array}
\right]=\left[
  \begin{array}{ccc}
A_{11}^\updownarrow \,\,\, A_{12}^\updownarrow\,\,\,A_{13}^\updownarrow\\
A_{21}^\updownarrow \,\,\, A_{22}^\updownarrow\,\,\,A_{23}^\updownarrow\\
A_{31}^\updownarrow \,\,\,A_{32}^\updownarrow\,\,\,A_{33}^\updownarrow\\
  \end{array}
\right],\quad\,
$$
\begin{eqnarray}
\bar{\textbf{A}}^\updownarrow=|\downarrow \rangle_1\otimes|\uparrow  \rangle_2 =
\left[
  \begin{array}{ccc}
{\bar{\phi}}_x^+\\
{\bar{\phi}}_y^+\\
{\bar{\phi}}_z^+\\
  \end{array}
\right]
\left[
\begin{array}{ccc}
 {\bar{\phi}}_x^-\,\,
 {\bar{\phi}}_y^- \,\,
 {\bar{\phi}}_z^-
  \end{array}
  \right]=\left[
  \begin{array}{ccc}
\bar{\phi}_x^+\bar{\phi}_x^- \,\,\,\bar{\phi}_x^+\bar{\phi}_y^-\,\,\,\bar{\phi}_x^+\bar{\phi}_z^-\\
\bar{\phi}_y^+\bar{\phi}_x^- \,\,\,\bar{\phi}_y^+\bar{\phi}_y^-\,\,\,\bar{\phi}_y^+\bar{\phi}_z^- \\
\bar{\phi}_z^+\bar{\phi}_x^- \,\,\,\bar{\phi}_z^+\bar{\phi}_y^-\,\,\,\bar{\phi}_z^+\bar{\phi}_z^-\\
  \end{array}
\right]=\left[
  \begin{array}{ccc}
 \bar{A}_{11}^\updownarrow\,\,\,\bar{A}_{12}^\updownarrow\,\,\,\bar{A}_{13}^\updownarrow\\
\bar{A}_{21}^\updownarrow\,\,\,\bar{A}_{22}^\updownarrow\,\,\,\bar{A}_{23}^\updownarrow\\
\bar{A}_{31}^\updownarrow\,\,\,\bar{A}_{32}^\updownarrow\,\,\,\bar{A}_{33}^\updownarrow\\
  \end{array}
\right],\nonumber\\
\label{A.01ab}
\end{eqnarray}
whereas the direct tensor product between vectors with parallel spins can be represented as:
\begin{eqnarray}
\textbf{A}^\upuparrows=|\uparrow\rangle_1\otimes|\uparrow\rangle_2 =
\left[
  \begin{array}{ccc}
{\phi}_x^+\\
{\phi}_y^+\\
{\phi}_z^+\\
  \end{array}
\right]
\left[
\begin{array}{ccc}
 \bar{{\phi}}_x^-\,\,
 \bar{{\phi}}_y^- \,\,
\bar{{\phi}}_z^-
  \end{array}
\right]=\left[
  \begin{array}{ccc}
{\phi}_x^+\bar{{\phi}}_x^- \,\,\, {\phi}_x^+\bar{{\phi}}_y^-\,\,\,{\phi}_x^+\bar{{\phi}}_z^-\\
{\phi}_y^+\bar{{\phi}}_x^- \,\,\, {\phi}_y^+\bar{{\phi}}_y^-\,\,\,{\phi}_y^+\bar{{\phi}}_z^-\\
{\phi}_z^+\bar{{\phi}}_x^- \,\,\, {\phi}_z^+\bar{{\phi}}_y^-\,\,\,{\phi}_z^+\bar{{\phi}}_z^-\\
  \end{array}
\right]=\left[
  \begin{array}{ccc}
A_{11}^\upuparrows \,\,\, A_{12}^\upuparrows\,\,\,A_{13}^\upuparrows\\
A_{21}^\upuparrows\,\,\, A_{22}^\upuparrows\,\,\,A_{23}^\upuparrows\\
A_{31}^\upuparrows\,\,\,A_{32}^\upuparrows\,\,\,A_{33}^\upuparrows\\
  \end{array}
\right],
\nonumber\\
\textbf{A}^\downdownarrows=|\downarrow\rangle_1\otimes|\downarrow\rangle_2 =
\left[
  \begin{array}{ccc}
\bar{{\phi}}_x^+\\
\bar{{\phi}}_y^+\\
\bar{{\phi}}_z^+\\
  \end{array}
\right]
\left[
\begin{array}{ccc}
{\phi}_x^-\,\,
{\phi}_y^- \,\,
{\phi}_z^-
  \end{array}
\right]=\left[
  \begin{array}{ccc}
\bar{{\phi}}_x^+{\phi}_x^- \,\,\,\bar{\phi}_x^+{\phi}_y^-\,\,\,\bar{\phi}_x^+{\phi}_z^-\\
\bar{\phi}_y^+{\phi}_x^- \,\,\,\bar{\phi}_y^+{\phi}_y^-\,\,\,\bar{\phi}_y^+{\phi}_z^- \\
\bar{\phi}_z^+{\phi}_x^- \,\,\,\bar{\phi}_z^+{\phi}_y^-\,\,\,\bar{\phi}_z^+{\phi}_z^-\\
  \end{array}
\right]=\left[
  \begin{array}{ccc}
A_{11}^\downdownarrows\,\,\, A_{12}^\downdownarrows\,\,\,A_{13}^\downdownarrows\\
A_{21}^\downdownarrows\,\,\, A_{22}^\downdownarrows\,\,\,A_{23}^\downdownarrows\\
A_{31}^\downdownarrows\,\,\,A_{32}^\downdownarrows\,\,\,A_{33}^\downdownarrows\\
  \end{array}
\right],\nonumber\\
  \label{A.01b}
\end{eqnarray}
where $\textbf{A}^\updownarrow$ and $\textbf{A}^\upuparrows$
denote the third-rank matrices, while $\bar{\textbf{A}}^\updownarrow$ and
$\bar{\textbf{A}}^\upuparrows=\textbf{A}^\downdownarrows$
are their complex conjugate matrices.

\subsection{T\lowercase{he zero-spin particles and the scalar field}}

Now the main question is the question of the so-called \emph{quintessence}-
is it possible to form particle-like excitations in the form of some dynamic
scalar field \cite{Steinh}?

Using the first equation of the system (\ref{4.03b}), we can construct the
boson wave function with zero spin, \textbf{entangling two hions with opposite spin
projections}, presenting it in the form (see Appendix B):
\begin{eqnarray}
\bm\phi^{\updownarrow}_\mp(\textbf{r}_+,\textbf{r}_-)=\frac{1}{\sqrt{2}}
\bigl\{\textbf{A}^\updownarrow\mp\bar{\textbf{A}}^\updownarrow\}=\frac{1}{\sqrt{2}}\,
\textbf{B}_\mp=
 \frac{1}{\sqrt{2}}\left[
  \begin{array}{ccc}
B_{11}^\mp\,\,\,B_{12}^\mp\,\,\,B_{13}^\mp\\
B_{21}^\mp\,\,\,B_{22}^\mp\,\,\,B_{23}^\mp\\
B_{31}^\mp\,\,\,B_{32}^\mp\,\,\,B_{33}^\mp\\
  \end{array}
\right]=
\frac{1}{\sqrt{2}}
\left[
 \begin{array}{ccc}
B_{11}^\mp\quad 0\quad 0\,\,\,\\
0\quad B_{22}^\mp\,\,\,\, 0\,\\
\,\,0\quad\, 0\,\,\,\,\, B_{33}^\mp\\
  \end{array}
\right],\,\,\nonumber\\
\label{4.03bz}
\end{eqnarray}
where the matrix elements $B^\mp_{ij}=A_{ij}\mp\bar{A}_{ij}$ are calculated explicitly.
Taking into account the rule of localization of the wave function components on the
corresponding planes (see Fig. 2), we obtain:
\begin{widetext}
\begin{eqnarray}
B^-_{11}=2i\bigl[\phi^{+(i)}_x\phi^{-(r)}_x-\phi^{+(r)}_x\phi^{-(i)}_x\bigr],\qquad
B^+_{11}=2\bigl[\phi^{+(r)}_x\phi^{-(r)}_x+\phi^{+(i)}_x\phi^{-(i)}_x\bigr],
\nonumber\\
B^-_{22}=2i\bigl[\phi^{+(i)}_y\phi^{-(r)}_y-\phi^{+(r)}_y\phi^{-(i)}_y\bigr],\qquad
B^+_{22}=2\bigl[\phi^{+(r)}_y\phi^{-(r)}_y+\phi^{+(i)}_y\phi^{-(i)}_y\bigr],
\nonumber\\
B^-_{33}=2i\bigl[\phi^{+(i)}_z\phi^{-(r)}_z-\phi^{+(r)}_z\phi^{-(i)}_z\bigr],\qquad
B^+_{33}=2\bigl[\phi^{+(r)}_z\phi^{-(r)}_z+\phi^{+(i)}_z\phi^{-(i)}_z\bigr].
 \label{4.04}
\end{eqnarray}
\end{widetext}
Note that the matrix elements with the plus sign are zero $B^+_{ij}=0$, since
it is possible easy to show that the components of the corresponding wave functions
are localized on disjoint manifolds. Latter means that the wave state
$\bm\phi^{\updownarrow}_+(\textbf{r}_+,\textbf{r}_-)$ does not exist.  In other words,
in the case of the Minkowski space-time there is only one singlet state for zero-spin boson.

The quantum distribution of the scalar boson in the singlet state before the onset of the
relaxation process, ie, for $s=0$, can be represented as:
 \begin{eqnarray}
 \mathcal{{P}}^0(\{\bm\phi^+\},\{\bm\phi^-\})=
  \bigl|\bigr|\bm\phi_-^\updownarrow(\textbf{r}_+,\textbf{r}_-)
 \bigr|\bigr|^2 = \frac{1}{2}\, \textbf{C},
\label{4.03bz}
\end{eqnarray}
where
$$\textbf{C}=\textbf{B}\cdot\bar{\textbf{B}}=
\left[
 \begin{array}{ccc}
 C_{11}\quad 0\quad 0  \\
0\quad C_{22}\quad0\\
  0\quad 0\quad C_{33}\\
  \end{array}
\right],
$$
is a diagonal third rank matrix, the elements of which have the following form:
\begin{eqnarray}
C_{11}=B_{11}\bar{B}_{11},\quad C_{22}=B_{22}\bar{B}_{22},\quad
C_{33}=B_{33}\bar{B}_{33},
\quad
C_{13}=C_{12}=C_{23}=C_{32}=0.
\label{4.04}
\end{eqnarray}
Taking into account the fact that the spins of two vector states  $\bm\phi^+$ and
$\bm\phi^-$ are directed oppositely, and also considering features of spatial
localization these quasiparticles, we obtain the following expressions for the matrix elements:
 \begin{eqnarray}
C_{11}=4||\phi^{+(r)}_x||^2||\phi^{-(i)}_x||^2+4||\phi^{+(i)}_x||^2||\phi^{-(r)}_x||^2,
\nonumber\\
C_{22}=4||\phi^{+(r)}_y||^2||\phi^{-(i)}_y||^2+4||\phi^{+(i)}_y||^2||\phi^{-(r)}_y||^2,
\nonumber\\
C_{33}=4||\phi^{+(r)}_z||^2||\phi^{-(i)}_z||^2+4||\phi^{+(i)}_z||^2||\phi^{-(r)}_z||^2.
\label{4.04bzt}
\end{eqnarray}

Now let us consider how the density of the quantum distribution of a scalar boson
changes taking into account the random influence of the environment.

To study this problem, we will use the following system of complex stochastic matrix equations:
\begin{eqnarray}
\partial_{t}\bm{\breve{{\psi}}^{\pm}}(\textbf{r}_\pm,t)\mp \bigl(\textbf{S}\cdot \bm\nabla\bigr) \bm{\breve{{\psi}}^{\pm}}(\textbf{r}_\pm,t)=\bm\eta^\pm(s_\pm),
\label{4.02t}
\end{eqnarray}
and also the equations:
\begin{eqnarray}
\nabla\bm{\breve{{\psi}}^{\pm}}(\textbf{r}_\pm,t)=0,
\label{4.02k}
\end{eqnarray}
where $\textbf{r}_\pm$ denotes the radius-vector of corresponding \emph{hion},
in addition, the complex  generators $\bm\eta^\pm(s_\pm)=
(\eta_x^\pm,\eta_y^\pm,\eta_z^\pm)$ describing random fluctuations of charges and
currents, which continuously arise in  4$D$-interval $ds^2_\pm=c^2dt^2-dx^2_\pm-dy^2_\pm-dz^2_\pm$.

For further studies, the system of equations (\ref{4.02t}), it is
useful to write in the matrix form:
$$
\left[
\begin{array}{ccc}
ict & -z_+ & y_+\\
z_+ & ict & -x_+ \\
-y_+& x_+ & ict \\
\end{array}
\right]\cdot
\left[
\begin{array}{ccc}
 \dot{\breve{{\psi}}}_x^+\\
\dot{\breve{{\psi}}}_y^+ \\
\dot{\breve{{\psi}}}_z^+\\
  \end{array}
\right]=s_+\left[
  \begin{array}{ccc}
  \eta_x^+ \\
  \eta_y^+ \\
    \eta_z^+\\
  \end{array}
\right],\quad
$$
and, respectively,
\begin{eqnarray}
\label{4.01a}
\left[
  \begin{array}{ccc}
  ict & z_- &- y_-\\
 -z_- & ict & x_- \\
  y_-& -x_- & ict \\
  \end{array}
\right]\cdot
\left[
\begin{array}{ccc}
\dot{\breve{{\psi}}}_x^-\\
\dot{\breve{{\psi}}}_y^- \\
\dot{\breve{{\psi}}}_z^-\\
  \end{array}
\right]=s_-\left[
  \begin{array}{ccc}
\eta_x^- \\
\eta_y^- \\
\eta_z^-\\
  \end{array}
\right],
\end{eqnarray}
where $ \dot{\breve{{\psi}}}_\sigma^\varsigma=\partial
\breve{{\psi}}_\sigma^\varsigma/\partial s_\varsigma$  and $\varsigma=\pm.$

As in the case of one \emph{hion} (see (\ref{4.s04})-(\ref{4.s05})),
the system of SDE (\ref{4.01a}) can be reduced to the canonical form:
\begin{eqnarray}
\dot{\breve{{\psi}}}^{\pm}_\sigma(s_\pm;\textbf{r}_\pm,t)=\bigl\{b^\pm_\sigma(\textbf{r}_\pm,t)+
d^\pm_\sigma(\textbf{r}_\pm,t)\bigr\}\bar{s}^{-1}_\pm\eta^-(s_\pm), \quad
\label{4.02a}
\end{eqnarray}
where $\bar{s}_+=s_+/a_p$ and $\bar{s}_-=s_-/a_p$, in addition, the following notations are made:
\begin{eqnarray}
b^\varsigma_x=\frac{z_\varsigma-y_\varsigma}{a_p},\qquad
d_x^{\,\varsigma}=\frac{c^2t^2-x^2_\varsigma-x_\varsigma y_\varsigma-x_\varsigma z_\varsigma}{a_pct},
\nonumber\\
b^\varsigma_y=\frac{x_\varsigma-z_\varsigma}{a_p},\qquad
d^{\,\varsigma}_y=\frac{c^2t^2-y^2_\varsigma-x_\varsigma y_\varsigma-y_\varsigma z_\varsigma}{a_pct},
\nonumber\\
 b^\varsigma_z=\frac{y_\varsigma-x_\varsigma}{a_p},
\qquad
d^{\,\varsigma}_z=\frac{c^2t^2-z^2_\varsigma-x_\varsigma z_\varsigma-y_\varsigma z_\varsigma}{a_pct}.
\label{4.02b}
\end{eqnarray}
Below in the equations (\ref{4.02a})-(\ref{4.02b}), we will assumed that
the following relations are satisfied:
$$
\eta_x^\varsigma=\eta_y^\varsigma=\eta_z^\varsigma=\eta,\qquad  d\bar{s}_+=d\bar{s}_-=d\bar{s},
$$
which is quite natural.

As in the case of a single \emph{hion}, we will assume that the random
 generator $\eta(s)$ satisfies the correlation properties of white noise (see Eqs. (\ref{4.s06})).

The joint probability distribution for a scalar boson can be represented as (see \cite{AshG}):
 \begin{eqnarray}
\mathcal{P}(\{\breve{\psi}\},\bar{s};\textbf{r}_+,\textbf{r}_-,t)=\prod_{\varsigma,
\sigma }\bigl\langle\delta\bigl(\breve{\psi}_\sigma^\varsigma(\bar{s};\textbf{r}_\varsigma,t)-
\breve{\phi}_{\sigma}^\varsigma\bigr)\bigr\rangle,
\label{4.02abc}
\end{eqnarray}
where  $\{{\breve{\psi}}\}=({\breve{\psi}}^+_x,...,{\breve{\psi}}^-_z)$ denotes a set of
fluctuating vacuum fields and $\breve{\phi}_{\sigma}^\varsigma=
\breve{\psi}_{\sigma}^\varsigma(\bar{s};\textbf{r},t)|_{s\to0}$.
In the representation (\ref{4.02abc}) the function
$\delta(\breve{\psi}_\sigma^\varsigma(s;\textbf{r}_\varsigma,t)\breve{\phi}_{\sigma}^\varsigma)$
denotes the Dirac delta function generalized on a 6$D$-Hilbert space.

Using the SDE system (\ref{4.02a}), for the conditional probability
 describing the relaxation of the singlet state, we can obtain the
following partial differential equation of the second order (see \cite{Ashg1}):
\begin{eqnarray}
\Bigl\{\frac{\partial }{\partial\bar{s}}-\frac{1}{2} \sum_{\varsigma,\sigma }
\varepsilon_{1\sigma}^\varsigma(\textbf{r}_\varsigma,t) \frac{\partial^2 }{\partial{\breve{\psi}}_\sigma^\varsigma\partial\bar{\breve{\psi}}_\sigma^\varsigma}
\Bigr\}\mathcal{P}=0,
\label{4.03}
\end{eqnarray}
where $\bar{\breve{\psi}}_\sigma^\varsigma$ denotes the complex conjugate of the function
$\breve{\psi}_\sigma^\varsigma$ and $\varepsilon_{1\sigma}^\varsigma(\textbf{r}_\varsigma,t)=
\varepsilon_1 [b_\sigma^\varsigma(\textbf{r}_\varsigma,t)+
d_\sigma^\varsigma(\textbf{r}_\varsigma,t)]^{2}$.

For further analytical calculations, it is convenient to represent the general
solution of the equation (\ref{4.03}) in the integral form:
\begin{eqnarray}
\mathcal{P}(\{\breve{\bm\psi}\},\bar{s}\,;\textbf{r}_+,\textbf{r}_-,t)=\int_{\Xi^6}
R(\{\breve{\bm\phi}^+\},\{\breve{\bm\phi}^-\})\prod_{\varsigma,\sigma}
\exp\biggl\{-\frac{(\breve{\phi}_{\sigma}^\varsigma-{\breve{\psi}}_\sigma^\varsigma)
(\bar{\breve{\phi}}_{\sigma}^\varsigma
-\bar{{\breve{\psi}}}_\sigma^\varsigma) }{2\bar{s}\varepsilon_{1\sigma}^\varsigma}\biggr\}
\frac{d \breve{\bm\phi}_{\sigma}^\varsigma}{ \sqrt{2\pi \bar{s}\varepsilon_{1\sigma}^\varsigma}},\quad
\label{4.03a}
\end{eqnarray}
where $d\breve{\bm\phi}_{\sigma}^\varsigma=d\breve{\phi}_{\sigma}^{\varsigma(r)}
d\breve{\phi}_{\sigma}^{\varsigma(i)}$,  in addition, as in the case of a single \emph{hion}, we assume that  $R(\{\breve{\bm\phi}^+\},\{\breve{\bm\phi}^-\})=\mathcal{{P}}^0(\{\bm\phi^+\},\{\bm\phi^-\})$
is the initial distribution of the scalar boson before the relaxation begins.
It is obvious that integration over the space-time, ie, by the spectrum, in
accordance with the ergodic hypothesis, is equivalent to integration
over the full 12$D$ space.

Substituting $(\ref{4.03bz})-(\ref{4.04bzt})$ into $(\ref{4.03a})$ and integrating by
variables $ \breve{\bm\phi}_{\sigma}^\varsigma$ within $[{\breve{\psi}}_\sigma^{\varsigma(\varpi)},
\phi_{\sigma}^{\varsigma(\varpi)}]$, we get:
\begin{eqnarray}
 \mathcal{P}(\{\breve{\bm\psi}\},\bar{s}\,;\textbf{r}_+,\textbf{r}_-,t)
= \frac{1}{2}\left[
 \begin{array}{ccc}
\overline{C}_{11}\,\,\,\,0\quad 0 \\
0\quad \overline{C}_{22}\,\,\,  0 \\
 \,\,0\quad\, 0\,\,\,\,  \overline{C}_{33}\\
 \end{array}
\right].
\label{4.05w}
\end{eqnarray}
Recall that $\overline{C}_{\sigma\sigma}$ denotes the mean value:
$$
\overline{C}_{\sigma\sigma}=4\Bigl\{\bigl\langle||\phi^{+(r)}_\sigma||^2\bigr\rangle
\bigl\langle||\phi^{-(i)}_\sigma||^2\bigr\rangle +
\bigl\langle||\phi^{+(i)}_\sigma||^2\bigr\rangle
\bigl\langle||\phi^{-(r)}_\sigma||^2\bigr\rangle\Bigr\},
$$
where
$$
\bigl\langle||\phi^{\varsigma(\varpi)}_\sigma||^2\bigr\rangle=
||\phi^{\varsigma(\varpi)}_\sigma||^2F_\sigma^{\varsigma(\varpi)}
\bigl(\phi^{\varsigma(\varpi)}_\sigma,\breve{\psi}^{\varsigma(\varpi)}_\sigma\bigr),
$$
in addition, performing similar calculations, as in the case of a single \emph{hion}
(see (\ref{4.s09})-(\ref{4.s14})), we can get:
\begin{eqnarray}
F_\sigma^{\varsigma(\varpi)}= \frac{1}{2}
\biggl\{1+\mathrm{erf}\biggl[\frac{\phi^{\varsigma(\varpi)}_\sigma-
\breve{\psi}^{\varsigma(\varpi)}_\sigma}{\sqrt{2\bar{s}\epsilon_{1\sigma}^\varsigma}}\biggr]\biggr\}.
\label{4.t01}
\end{eqnarray}

Note that, using analogous arguments, we can construct the wave function of entangled two
\emph{hions} with consideration of its relaxation in a random environment. Also,
performing calculations similarly to the case of one \emph{hion}, one can
see that the deformation of the wave state of one scalar boson under the influence of
 a random environment does not lead to a violation of the law of conservation of full probability.

Thus, we have shown that, as a result of the multi-scale evolution of QVF, a scalar
field is formed, as a sort of Bose-Einstein condensate of massless scalar bosons.
However, such a condensate differs significantly from a conventional substance,
since it consists of massless particles that have a large Compton wavelength that
can not thicken unlimitedly and to form large-scale structures such as stars, planets, etc.
In other words, the described substance meets all the characteristics of the
\emph{quintessence} requirements and, accordingly, it can be  asserted that the
\emph{quintessence  hypothesis} is theoretically proved.

\subsection{T\lowercase{riplet state of two \emph{hions} and the vector field}}
The wave function of the triplet state formed by entangling two \emph{hions}
with \textbf{parallel} projections of the spin can be represented as:
\begin{eqnarray}
\bm\phi^{\upuparrows}_\mp(\textbf{r}_+,\textbf{r}_-)=\frac{1}{\sqrt{2}}
\bigl\{\textbf{A}^\upuparrows\mp \textbf{A}^\downdownarrows\}=
\frac{1}{\sqrt{2}}\,\textbf{D}_\mp=
 \frac{1}{\sqrt{2}}\left[
  \begin{array}{ccc}
D_{11}^\mp\,\,\,D_{12}^\mp\,\,\,D_{13}^\mp\\
D_{21}^\mp\,\,\,D_{22}^\mp\,\,\,D_{23}^\mp\\
D_{31}^\mp\,\,\,D_{32}^\mp\,\,\,D_{33}^\mp\\
  \end{array}
\right]=
\frac{1}{\sqrt{2}}
\left[
 \begin{array}{ccc}
D_{11}^\mp\quad 0\quad 0\,\,\,\\
0\quad D_{22}^\mp\,\,\,\, 0\,\\
\,\,0\quad\, 0\,\,\,\,\,D_{33}^\mp\\
  \end{array}
\right],\,\,\nonumber\\
\label{4.04bz}
\end{eqnarray}
where, as shown by simple calculations, the following equalities hold:
\begin{equation}
D^\mp_{11}=B^\mp_{11},\qquad D^\mp_{22}=B^\mp_{22},\qquad D^\mp_{33}=B^\mp_{33},
\label{4.04z}
\end{equation}
From these equalities it follows that there is only one triplet
state, which described by the matrix of the third rank
$\bm\phi^{\upuparrows}_-(\textbf{r}_+,\textbf{r}_-)$.

The relaxation of the triplet state (\ref{4.04bz}) can be taken into account
using a similar construction, as in the case of the singlet state (\ref{4.03a}).
In this case, however, the following substitutions
$(x_-\to -x_-,\quad y_-\to -y_-,\quad z_-\to -z_-)$ must be made in the expressions
(\ref{4.02a})-(\ref{4.02b}), which will be equivalent to transforming the singlet
state into the triplet state. Note that these replacements in the expression
(\ref{4.03a}) changes the power of fluctuations $\varepsilon_\sigma^\varsigma$.

\section{C\lowercase{onclusion}}
Although a fundamental scalar field has not yet been observed experimentally,
it is generally accepted that such fields play a key role in the construction
of modern theoretical physics of elementary particles. There are a few important
hypothetical scalar fields, for example the Higgs field for the Standard Model,
the \emph{dark energy}-\emph{quintessence} for a theory of the quantum vacuum, etc.
Note that the presence of each of them is necessary for the complete classification
of the theory of fundamental fields, including new physical theories, such as,
for example, String Theory.
Recall that despite the great progress in the representations of modern particle theory
within the framework of the SM, it  does not give a clear explanation of a number of
fundamental questions of the modern physics, such as \emph{What is dark energy and dark matter}?
or \emph{What happened to the antimatter after the big bang}?  and so on.
As modern astrophysical observations show, not less than 74 percent of the energy of
the universe is associated with a substance called \emph{dark energy}, which has no
mass and whose properties are not sufficiently studied and understood. Based on
many considerations, it was obvious to assume that this substance must be related
to a quantum vacuum or simply be QV itself.

As it is known, in the modern understanding of what is called the vacuum state or
the quantum vacuum, it is  by no means a \emph{simple empty space}. Recall that
in the vacuum state, electromagnetic waves and particles continuously appear and
disappear, so that on the average their value is zero.  It would be reasonable to
think that these fluctuating or flickering fields are born as a result of spontaneous
decays of quasi-stable scalar massless particles, very inert to any external influences.

The main purpose of this work was the theoretical justification for the possibility
of forming a scalar field consisting of uncharged massless zero-spin particles.
The  developed approach is formally similar to the Parisi-Wu \cite{Parisi-Wu, Dam}
stochastic quantization, however it also has significant differences.
In particular, as in the case of the Parisi-Wu stochastic quantization, when considering
Euclidean quantum field theory, we consider the stochastic Yang-Mills equations as the
basic equations. However, unlike the Parisi-Wu concept, we believe that the nature of
stochasticity is multi-scale and, therefore, the equilibrium limit of the statistical
system associated with a heat reservoir is not one, but many. In other words, there
are many quasi-equilibrium states between which spontaneous transitions occur,
but at the same time the dynamical equilibrium between these states is conserved.
Another significant difference between the developed representation and the
Parisi-Wu theory is that the analogy with classical statistical physics is
not used for determination the stationary distribution of a random process.
The latter circumstance allows us to avoid a number of inaccuracies inherent
in standard representations, which, in our opinion, makes it difficult to
study such a specific substance as QV.

In this article, we have considered the simplest case when the self-action
terms are absent in the Yang-Mills stochastic equations, that is, $\mathfrak{g}=0$.
This means that considered QVF  are Abelian fields, which
satisfy the gauge group symmetry $SU(2)\times U(1)$. We quantized
the classical stochastic vector field (see quantization conditions (\ref{3.04ak}))
and proved that on the main relaxation scale $(\tau_0,\bm\varepsilon_0^a)$, in the
Hilbert space, in the limit of statistical equilibrium, a discrete set
of stationary solutions (\ref{3.03tz}) arises, describing a massless spin-1 Bose
particle (named \emph{hion}). It is important to note that these solutions,
combining relativism with quantum mechanics together (see Eqs (\ref{2.0k2b})),
are as close as possible to the concept of  2$D$ quantum string theory.
It is shown that \emph{hion} is characterized by three quantum numbers, where
the \emph{ground state} is characterized by the highest frequency. As shown in our study,
 the \emph{hion} in the $3D$ space is localized on the complex 2$D$
surface consisting of three perpendicular planes (see Fig. 2).

We have proved that on the second scale of relaxation  $(\tau_1,\bm\varepsilon_1)$,
\emph{hion} becomes quasi-particle, which can make spontaneous transitions
to other mass and massless states. Note that  during spontaneous
transitions \emph{hion} the bosons $W_3$ and $B$ combine into two different
bosons, such as the photon $\gamma$ (electromagnetic interaction) and the massive
boson $Z^0$ (weak interaction). It is also shown that on the second relaxation scale,
two \emph{hions} with spin projections +1 and -1 can form a boson with zero spin. The
ensemble of spin-0 bosons forms a Bose-Einstein condensate, which is a scalar
field with all the necessary properties. In other words, the work is a theoretical
proof of the \emph{dark energy}-\emph{quintessence} hypothesis and, accordingly,
the stability of the QED vacuum in the infrared limit. Recall that in the infrared
limit of vacuum diagrams does not exist in theories with self-interacting massless
fields (QCD) or massless interacting particles (massless QED), if the theory is
renormalizable \cite{Savi,SCHARF}.

Note that a small part of the energy of a quantum vacuum is concentrated in
vector fields  consisting of  \emph{hions} and vector bosons with spin-2.
These fields on a large scales of $3D$ space can be represented as Heisenberg
spin glass, the total  polarization of which is zero.

A very important question related to the value of the parameter $a_p$
(see (\ref{3.05at})), which characterizes the spatial size of the \emph{hion},
remains open within the framework of the developed representation. Apparently,
we can get a clear answer to this question by conducting a series of experiments.
In particular, if the value of the constant $a_p$ will be substantially
different from the Planck length $l_P$, then it will be necessary to introduce
a new fundamental constant  defining the spatial size of the \emph{hion}.
We are convinced that if it will be impossible to conduct direct measurements
of constants $({\varepsilon}_0^1,{\varepsilon}_0^2,{\varepsilon}_0^3)$,
characterizing the fluctuations of quintessence, and, accordingly, formation
of \emph{hions}, then this can be done using of indirect measurements.  In
particular, we believe that interesting experiments on the detection of optical
force between controlled light waves, known as ``optical binding strength" (see
for example \cite{Mo}), are not related to the Casimir vacuum properties, as some
researchers are trying to explain. The magnitude of the optical force and its
diverse behavior do not allow us to hope for it. To explain these forces, most
likely, there must be a more ``fundamental vacuum" with more non-trivial properties
like as scalar field or dark energy. A different experiments aimed at non-invasive
measurements of various characteristics of biological organisms and systems with
fluctuating entropies also push us to the idea of the existence of quintessence,
with  its still unknown properties \cite{Ash3}. 

Thus, studies carried out within the framework of the symmetry group
$SU(2)\times U(1)$ allow us to speak of the properties of the QVF and,
accordingly, of the structure and properties of empty space-time down to the
distances $\sim10^{-15} m$, when strong interactions begin to dominate.
As preliminary studies show, the properties of space-time can change as a
result of the polarization of the vector component QV and, accordingly,
changes in the refractive indices of the vacuum due to orientational
effects of \emph{hion}s in external, even weak electromagnetic fields.
Obviously, in this case the photon-photon interaction and, accordingly,
the interaction of two light beams occurs according to another physical
mechanism, which differs from the mechanism described by the fourth-order
Feynman diagrams.

In other words, there is every reason to speak for the first time about
the real, rather than theoretical, possibility of implementing space-time
engineering.

Finally, as many researchers point out, beginning of the middle of the twentieth
century a new scientific-technological revolution began, based not on energy but
on information. In this regard, some researcher are rightly identified the
Universe with  giant quantum computer, which can  explain previously unexplained
features, most importantly, the co-existence in the universe of randomness and order,
and of simplicity and complexity (see, for example \cite{Lloyd}). In the light
of the above theoretical proofs, the quantum vacuum-quintessence or scalar field
is nothing more than a natural quantum computer with complex logic,
different from that currently being realized in practice.

\section{A\lowercase{ppendix}}
\subsection{}
Let us consider  the limit of statistical equilibrium
$\lim_{\,t \to \infty}\mathcal{P}^1(\zeta,t) =\bar{\mathcal{P}}^1(\zeta) $.
In this case, the partial differential equation  (\ref{2.3wv}) is transformed
into the ordinary differential equation of the form:
\begin{equation}
\biggl\{3\zeta+
\bigl(\zeta^2+\omega^2\bigl)\frac{d}{d\zeta}
+\frac{\varepsilon_0}{2}\frac{d^2}{d\zeta^2}\biggr\}\bar{\mathcal{P}}^1(\zeta)=0.
\label{1.a0}
\end{equation}

\textbf{Proposition.} \emph{If the function $\bar{\mathcal{P}}^1(\zeta)$ the solution of the
equation (\ref{1.a0}), then the integral:
\begin{equation}
\int_{-\infty}^{+\infty}\bar{\mathcal{P}}^1(\zeta)d\zeta<M,
\label{1.ta1}
\end{equation}
where $M=const>0$.}

We represent the solution of the equation (\ref{1.a0}) in the form:
 \begin{equation}
\bar{\mathcal{P}}^1(\zeta)=Z(\zeta){\mathcal{P}}^+(\zeta).
\label{1.a1}
\end{equation}
Let the function ${\mathcal{P}}^+(\zeta)$ satisfies the equation:
\begin{equation}
\biggl\{\bigl(\zeta^2+\omega^2\bigl)\frac{d}{d\zeta}
+\frac{\varepsilon_0}{2}\frac{d^2}{d\zeta^2}\biggr\}{\mathcal{P}}^+(\zeta)=0,
\label{1.a2}
\end{equation}
then it will look like:
$$
{\mathcal{P}}^+(\zeta)=C^+\int^{\zeta}_{-\infty} e^{-2\varepsilon_0^{-1}
 (\frac{1}{3}z^3+\omega^2z)}dz,
$$
where $C^+$ is the arbitrary constant.\\
 If we choose $C^+>0$ then  the function
${\mathcal{P}}^+(\zeta)$  will be positive on all axis $\zeta\in(-\infty,+\infty)$.

Substituting (\ref{1.a1}) into (\ref{1.a0}), we  get:
\begin{equation}
\biggl\{3\zeta+
\Bigl[\bigl(\zeta^2+\omega^2\bigl)+\varepsilon_0K(\zeta)\Bigr]
\frac{d}{d\zeta}+\frac{\varepsilon_0}{2}\frac{d^2}{d\zeta^2}\biggr\}Z(\zeta)=0,
\label{1.a3}
\end{equation}
where
$$
K(\zeta)=\frac{d{\ln\mathcal{P}}^+(\zeta)}{d\zeta}=
e^{-2\varepsilon_0^{-1}(\frac{1}{3}\zeta^3+\omega^2\zeta)}\biggl/\int^{\zeta}_{-\infty}
 e^{-2\varepsilon_0^{-1}(\frac{1}{3}z^3+\omega^2z)}dz.
$$
As follows from the analysis of the coefficient $K(\zeta)$,
on the axis $\zeta\in(-\infty,+\infty)$, it has types of uncertainties $\frac{0}{0}$
or $\frac{\infty}{\infty}$. Applying the L'H\^{o}pital's rule, for the coefficient
near the critical points $\zeta_i$, where uncertainties appear, we obtain the expression:
\begin{equation}
\lim_{\zeta\to \zeta_i} K(\zeta)=\frac{\frac{d}{d\zeta}\Bigl(e^{-2\varepsilon_0^{-1} (\frac{1}{3}\zeta^3+\omega^2\zeta)}\Bigr)}{\frac{d}{d\zeta}\Bigl(\int^{\zeta}_{-\infty} e^{-2\varepsilon_0^{-1}
 (\frac{1}{3}z^3+\omega^2z)}dz\Bigr)}=-2\frac{\zeta^2+\omega^2}{\varepsilon_0}.
\label{1.0a4}
\end{equation}
With consideration (\ref{1.0a4}) the equation (\ref{1.a3}) can be written in the form:
\begin{equation}
\biggl\{3\zeta-
 \bigl(\zeta^2+\omega^2\bigl)\frac{d}{d\zeta}+\frac{\varepsilon_0}{2}
 \frac{d^2}{d\zeta^2}\biggr\}Z(\zeta)=0.
\label{1.a4}
\end{equation}
In asymptotic domains, i.e. when $|\zeta|\gg1$ in the equation (\ref{1.a4}), we can
neglect the term $3\zeta$ and to obtain  the following solution:
 \begin{equation}
Z(\zeta)\sim
\int^\zeta_{-\infty}e^{2\varepsilon_0^{-1}(\frac{1}{3}z^3+\omega^2z)}dz>0,
\qquad |\zeta|\gg1,
\label{1.a5}
\end{equation}
whereas the asymptotic solution of the equation (\ref{1.a0}), respectively, is:
 \begin{eqnarray}
\bar{\mathcal{P}}^1(\zeta)\sim C^+\biggl(\int^{\zeta}_{-\infty} e^{-2\varepsilon_0^{-1}
(\frac{1}{3}z^3+\omega^2z)}dz\biggr)
\biggl(\int^\zeta_{-\infty} e^{2\epsilon_0^{-1}(\frac{1}{3}y^3+\omega^2y)}dy\biggr)>0,
\qquad  |\zeta|\gg1.
\label{1.a6}
\end{eqnarray}
Again using the L'H\^{o}pital's rule, it can be proved that:
$$
\lim_{|\zeta|\to \infty}\biggl|\frac{\acute{Z}(\zeta)}{Z(\zeta)}\biggr|=\zeta^2+\omega^2\gg1,
\qquad \acute{Z}(\zeta)=\frac{d}{d\zeta} Z(\zeta),
$$
from which flow out the estimation:
$$
e^{- 2 \epsilon_0^{-1}
(\frac{1}{3}\zeta^3+\omega^2\zeta)}\gg \int^{\zeta}_{-\infty}e^{-2\varepsilon_0^{-1}
(\frac{1}{3}z^3+\omega^2z)}dz,\qquad |\zeta|\gg1,
$$
and, respectively;
\begin{equation}
\bar{\mathcal{P}}^1(\zeta)<C^+e^{-2\varepsilon_{0}^{-1}
(\frac{1}{3}\zeta^3+\omega^2\zeta)} \int^\zeta_{-\infty}
e^{2 \varepsilon_{0}^{-1}(\frac{1}{3}z^3+\omega^2z)}dz.
\label{1.a7}
\end{equation}

Now let us represent the integral (\ref{1.ta1}) in the form of a sum from three terms:
\begin{equation}
 \int_{-\infty}^{+\infty}\bar{\mathcal{P}}^1(\zeta)d\zeta=
\biggl\{\,\int_{-\infty}^{-L} + \int_{-L}^{+L}
  +\int^{+\infty}_{+L}\,\biggr\}\bar{\mathcal{P}}^1(\zeta)d\zeta,
\label{1.a8}
\end{equation}
where $L>>1.$

For the first and third integrals, we can write down the following obvious estimates
(see (\ref{2.32a})):
\begin{equation}
\int_{-\infty}^{-L}\bar{\mathcal{P}}^1(\zeta)d\zeta<
\int_{-\infty}^{-L}{\mathcal{P}}^0(\zeta)d\zeta<
\int_{-\infty}^{+\infty}{\mathcal{P}}^0(\zeta)d\zeta=1,
\label{1.a9}
\end{equation}
and, respectively:
\begin{equation}
\int^{+\infty}_{+L}\bar{\mathcal{P}}^1(\zeta)d\zeta<
\int^{+\infty}_{+L}{\mathcal{P}}^0(\zeta)d\zeta<
\int_{-\infty}^{+\infty}{\mathcal{P}}^0(\zeta)d\zeta=1.
\label{1.a10}
\end{equation}
As for the second term in (\ref{1.a8}):
 \begin{equation}
 \int_{-L}^{+L}\bar{\mathcal{P}}^1(\zeta)d\zeta<M^\ast,
\label{1.a11}
\end{equation}
then it is obviously converges $M^\ast<\infty$, since the function
$\bar{\mathcal{P}}^1(\zeta)$ is bounded, and the integration is performed
on a finite interval.

Thus taking into account the estimations (\ref{1.a9})-(\ref{1.a11})
we can approve that the integral (\ref{1.ta1}) converges, i.e. the  estimate:
$$
 \int_{-\infty}^{+\infty}\bar{\mathcal{P}}^1(\zeta)d\zeta<M,
$$
where $M=const<\infty,$ is \textbf{correct}.\\
\emph{ \textbf{The proposition is proved.}}

\subsection{}
Using the systems of equations (\ref{1.02k})-(\ref{1.02ta})  and carrying out a
similar calculations for vacuum fields consisting of particles with projections
of spin -1, we can obtain the following system of equations:
\begin{eqnarray}
 \bigl\{\square+ [i(c_{,y}-c_{,z})+c_{,t}c^{-1}]c^{-2}\partial_t\bigr\}\psi^{-}_x=0,
\nonumber\\
\bigl\{\square+[i(c_{,z}-c_{,x})+c_{,t}c^{-1}]c^{-2}\partial_t\bigr\}\psi^{-}_y=0,
\nonumber\\
\bigl\{\square+[i(c_{,x}-c_{,y})+c_{,t}c^{-1}]c^{-2}\partial_t\bigr\}\psi^{-}_z=0.
 \label{A.l}
\end{eqnarray}
In (\ref{A.l}) substituting the solutions of the equations in the form:
 \begin{equation}
 \psi_\sigma^-(\textbf{r},t)=\exp{\biggl\{\int_{-\infty}^t \zeta_\sigma(t')dt'
 \biggr\}}\phi_\sigma^-(\textbf{r}),\qquad
 \sigma=x,y,z,
\label{A.2}
\end{equation}
we obtain the following system of stationary equations:
\begin{eqnarray}
\Bigl\{\triangle-\Bigl[\Bigl(\frac{\xi(t)}{c}\Bigr)^2+\frac{r-i(y-z)}{cr^2}\,
 \zeta(t)\Bigr]\Bigr\}\phi_x^-(\textbf{r})=0,
\nonumber\\
\Bigl\{\triangle-\Bigl[\Bigl(\frac{\xi(t)}{c}\Bigr)^2+\frac{r-i(z-x)}{c r^2}\,
\zeta(t)\Bigr]\Bigr\}\phi_y^-(\textbf{r})=0,
\nonumber\\
\Bigl\{\triangle-\Bigl[\Bigl(\frac{\xi(t)}{c}\Bigr)^2+\frac{r-i(x-y)}{cr^2}\,
\zeta(t)\Bigr]\Bigr\}\phi_z^-(\textbf{r})=0.
 \label{A.2a}
\end{eqnarray}
Further, carrying out standard arguments, we find the following stationary equations:
\begin{eqnarray}
\Bigl\{\triangle -\Bigl[\Bigl(\frac{\omega}{c}\Bigr)^2+ \frac{r-i(y-z)}{r^2}\varrho(\omega)\Bigr]
\Bigr\}\phi_x^-(\textbf{r})=0,
\nonumber\\
\Bigl\{\triangle -\Bigl[\Bigl(\frac{\omega}{c}\Bigr)^2+\frac{r-i(z-x)}{r^2}\varrho(\omega)\Bigr]
\Bigr\}\phi_y^-(\textbf{r})=0,
\nonumber\\
\Bigl\{\triangle-\Bigl[\Bigl(\frac{\omega}{c}\Bigr)^2+\frac{r-i(x-y)}{r^2}\varrho(\omega)\Bigr]\Bigr\}
 \phi_z^-(\textbf{r})=0.
\label{A.2a}
\end{eqnarray}
The solution of the system of equations is conducted  in a similar way, as for fields
$\phi^+(\textbf{r})$ (see Eq.s (\ref{3.020})-(\ref{3.02b})). In particular,
calculations show that the components of the vector boson with the spin -1
projection are localized on a manifold consisting of the following set of planes;
$\bigl[\phi_x^{-(r)}(Y,-Z),\,\,
\phi_x^{-(i)}(-Y,Z)\bigr],\quad\bigl[\phi_y^{-(r)}(X,-Z),$ $\,\,\phi_x^{-(i)}(-X,Z)\bigr]$ and
$\bigl[\phi_z^{-(r)}(X,-Y),\,\,\phi_x^{-(i)}(-X,Y)\bigr]$, respectively.

\end{document}